\documentclass[dvips]{article}
\usepackage{supertabular,lscape,epsfig}
\usepackage{amssymb}
\usepackage{amsmath}

\DeclareSymbolFont{ppa}{OT1}{ppl}{m}{it}
\DeclareMathSymbol{\vv}{\mathalpha}{ppa}{'166}

\begin{document}

\newcommand{\dd}{\,{\rm d}}
\newcommand{\ie}{{\it i.e.},\,}
\newcommand{\etal}{{\it et al.\ }}
\newcommand{\eg}{{\it e.g.},\,}
\newcommand{\cf}{{\it cf.\ }}
\newcommand{\vs}{{\it vs.\ }}
\newcommand{\zdot}{\makebox[0pt][l]{.}}
\newcommand{\up}[1]{\ifmmode^{\rm #1}\else$^{\rm #1}$\fi}
\newcommand{\dn}[1]{\ifmmode_{\rm #1}\else$_{\rm #1}$\fi}
\newcommand{\upd}{\up{d}}
\newcommand{\uph}{\up{h}}
\newcommand{\upm}{\up{m}}
\newcommand{\ups}{\up{s}}
\newcommand{\arcd}{\ifmmode^{\circ}\else$^{\circ}$\fi}
\newcommand{\arcm}{\ifmmode{'}\else$'$\fi}
\newcommand{\arcs}{\ifmmode{''}\else$''$\fi}
\newcommand{\MS}{{\rm M}\ifmmode_{\odot}\else$_{\odot}$\fi}
\newcommand{\RS}{{\rm R}\ifmmode_{\odot}\else$_{\odot}$\fi}
\newcommand{\LS}{{\rm L}\ifmmode_{\odot}\else$_{\odot}$\fi}

\newcommand{\Abstract}[2]{{\footnotesize\begin{center}ABSTRACT\end{center}
\vspace{1mm}\par#1\par
\noindent
{~}{\it #2}}}

\newcommand{\TabCap}[2]{\begin{center}\parbox[t]{#1}{\begin{center}
  \small {\spaceskip 2pt plus 1pt minus 1pt T a b l e}
  \refstepcounter{table}\thetable \\[2mm]
  \footnotesize #2 \end{center}}\end{center}}

\newcommand{\TableSep}[2]{\begin{table}[p]\vspace{#1}
\TabCap{#2}\end{table}}

\newcommand{\FigCap}[1]{\footnotesize\par\noindent Fig.\  %
  \refstepcounter{figure}\thefigure. #1\par}

\newcommand{\TableFont}{\footnotesize}
\newcommand{\TableFontIt}{\ttit}
\newcommand{\SetTableFont}[1]{\renewcommand{\TableFont}{#1}}

\newcommand{\MakeTable}[4]{\begin{table}[htb]\TabCap{#2}{#3}
  \begin{center} \TableFont \begin{tabular}{#1} #4 
  \end{tabular}\end{center}\end{table}}

\newcommand{\MakeTableSep}[4]{\begin{table}[p]\TabCap{#2}{#3}
  \begin{center} \TableFont \begin{tabular}{#1} #4 
  \end{tabular}\end{center}\end{table}}
\newcommand{\TabCapp}[2]{\begin{center}\parbox[t]{#1}{\centerline{
  \small {\spaceskip 2pt plus 1pt minus 1pt T a b l e}
  \refstepcounter{table}\thetable}
  \vskip2mm
  \centerline{\footnotesize #2}}
  \vskip3mm
\end{center}}

\newcommand{\MakeTableSepp}[4]{\begin{table}[p]\TabCapp{#2}{#3}\vspace*{-.7cm}
  \begin{center} \TableFont \begin{tabular}{#1} #4 
  \end{tabular}\end{center}\end{table}}

\newfont{\bb}{ptmbi8t at 12pt}
\newfont{\bbb}{cmbxti10}
\newfont{\bbbb}{cmbxti10 at 9pt}
\newcommand{\uprule}{\rule{0pt}{2.5ex}}
\newcommand{\douprule}{\rule[-2ex]{0pt}{4.5ex}}
\newcommand{\dorule}{\rule[-2ex]{0pt}{2ex}}
\def\thefootnote{\fnsymbol{footnote}}

\newenvironment{references}%
{
\footnotesize \frenchspacing
\renewcommand{\thesection}{}
\renewcommand{\in}{{\rm in }}
\renewcommand{\AA}{Astron.\ Astrophys.}
\newcommand{\AAS}{Astron.~Astrophys.~Suppl.~Ser.}
\newcommand{\ApJ}{Astrophys.\ J.}
\newcommand{\ApJS}{Astrophys.\ J.~Suppl.~Ser.}
\newcommand{\ApJL}{Astrophys.\ J.~Letters}
\newcommand{\AJ}{Astron.\ J.}
\newcommand{\IBVS}{IBVS}
\newcommand{\PASP}{P.A.S.P.}
\newcommand{\Acta}{Acta Astron.}
\newcommand{\MNRAS}{MNRAS}
\renewcommand{\and}{{\rm and }}
\section{{\rm REFERENCES}}
\sloppy \hyphenpenalty10000
\begin{list}{}{\leftmargin1cm\listparindent-1cm
\itemindent\listparindent\parsep0pt\itemsep0pt}}%
{\end{list}\vspace{2mm}}

\def\TYLDA{~}
\newlength{\DW}
\settowidth{\DW}{0}
\newcommand{\dw}{\hspace{\DW}}

\newcommand{\refitem}[5]{\item[]{#1} #2%
\def\REFARG{#3}\ifx\REFARG\TYLDA\else, {\it#3}\fi
\def\REFARG{#4}\ifx\REFARG\TYLDA\else, {\bf#4}\fi
\def\REFARG{#5}\ifx\REFARG\TYLDA\else, {#5}\fi.}

\newcommand{\Section}[1]{\section{\hskip-6mm.\hskip3mm#1}}
\newcommand{\Subsection}[1]{\subsection{#1}}
\newcommand{\Acknow}[1]{\par\vspace{5mm}{\bf Acknowledgements.} #1}
\pagestyle{myheadings}

\newcommand{\xrule}{\rule{0pt}{2.5ex}}
\newcommand{\xxrule}{\rule[-1.8ex]{0pt}{4.5ex}}
\def\thefootnote{\fnsymbol{footnote}}
\begin{center}
{\Large\bf The Optical Gravitational Lensing Experiment.\\
\vskip3pt
Catalog of RR Lyrae Stars\\
\vskip6pt
from the Small Magellanic Cloud\footnote{Based on
observations obtained with the 1.3~m Warsaw telescope at the Las Campanas
Observatory of the Carnegie Institution of Washington.}}
\vskip1.2cm
{\bf I.~~S~o~s~z~y~\'n~s~k~i$^{1,2}$,~~A.~~U~d~a~l~s~k~i$^1$,~~M.~~S~z~y~m~a~{\'n}~s~k~i$^1$,\\
M.~~K~u~b~i~a~k$^1$,~~G.~~P~i~e~t~r~z~y~\'n~s~k~i$^{1,3}$,~~
P.~~W~o~\'z~n~i~a~k$^4$,\\ K.~\.Z~e~b~r~u~\'n$^1$,
~~O.~~S~z~e~w~c~z~y~k$^1$~ and ~\L.~~W~y~r~z~y~k~o~w~s~k~i$^1$}
\vskip8mm
{$^1$Warsaw University Observatory, Al.~Ujazdowskie~4, 00-478~Warszawa,
Poland\\
e-mail: (soszynsk,udalski,msz,mk,pietrzyn,zebrun,szewczyk,wyrzykow)@astrouw.edu.pl\\
$^2$ Princeton University Observatory, Princeton, NJ 08544-1001 USA\\
$^3$ Universidad de Concepci{\'o}n, Departamento de Fisica,
Casilla 160--C, Concepci{\'o}n, Chile\\
$^4$ Los Alamos National Laboratory, MS-D436, Los Alamos, NM 87545 USA\\
e-mail: wozniak@lanl.gov}
\end{center}

\vskip1.3cm
\Abstract{We present the catalog of RR~Lyrae stars from 2.4 square degrees 
of central parts of the Small Magellanic Cloud (SMC). The photometric data 
were collected during four years of the OGLE-II microlensing survey. 
Photometry of each star was obtained using the Difference Image Analysis (DIA) 
method. The catalog contains 571 objects, including 458~RRab, 56~RRc 
variables, and 57 double mode RR~Lyr stars (RRd). Additionally we attach
a~list of variables with periods between 0.18--0.26 days, which are probably 
$\delta$~Sct stars. Period, {\it BVI} photometry, astrometry, amplitude, and 
parameters of the Fourier decomposition of the {\it I}-band light curve are 
provided for each object. We also present the Petersen diagram for double mode 
pulsators. 

We found that the SMC RR~Lyr stars are fairly uniformly distributed over the 
studied area, with no clear correlation to any location. The most preferred 
periods for RRab and RRc stars are 0.589 and 0.357 days, respectively. We 
noticed significant excess of stars with periods of about 0.29 days, which 
might be second-overtone RR~Lyr stars (RRe). The mean extinction free 
magnitudes derived for RRab stars are 18.97, 19.45 and 19.73 mag for the $I$, 
$V$ and {\it B}-band, respectively. 

All presented data, including individual {\it BVI} observations, are available 
from the OGLE {\sc Internet} archive.}{}

\Section{Introduction}
\vskip13pt
The Magellanic Clouds provide an ideal opportunity to study in detail the 
structure and evolution of stars. Rich populations of stars approximately at 
the same, relatively small distance with small interstellar reddening make the 
Large and Small Magellanic Clouds very important targets for observing 
surveys. 

RR~Lyr stars were first discovered in the SMC near the cluster NGC121 by 
Thackeray (1951). The stars were more than a~magnitude fainter than expected, 
what was a~crucial confirmation of the major revision of the extragalactic 
distance scale proposed by Baade (1952). The first systematic search for field 
RR~Lyr variables in the SMC was conducted by Graham (1975). During that survey 
76 RR~Lyr stars were discovered in a~${1\arcd\times1\zdot\arcd3}$ outlying 
field centered on the cluster NGC121. Smith \etal (1992) used {\it B}-band 
photographic photometry of a~field in the northeast arm of the SMC to 
identify additional 22 probable RR~Lyr variables. The same outlying region of 
the SMC was observed by Sharpee \etal (2002). They presented {\it V}-band and 
{\it B}-band CCD photometry of a~few RR~Lyr stars. 

Walker (1989) surveyed five SMC clusters for RR~Lyr stars. Four already known 
RR~Lyr variables in the NGC121 were rediscovered, but no such stars were found 
in the other clusters. Because the age of NGC121 was estimated to be 
$12\pm2$~Gyr (Stryker \etal 1985), and Lindsay~1, the next oldest SMC cluster, 
$10\pm2$~Gyr (Olszewski \etal 1987), Walker concluded that the minimum age of 
RR~Lyr stars is about 11~Gyr. 

In 1990s the number of known variable stars in the Magellanic Clouds 
dramatically increased, when the large microlensing searches began regular 
photometric monitoring of both galaxies (\eg MACHO -- Alcock \etal 1993, EROS 
-- Aubourg \etal 1993). Natural by-product of the microlensing surveys are 
huge databases with precise photometry of millions of stars. 

The SMC was also included to the list of targets of the second phase of the 
Optical Gravitational Lensing Experiment (OGLE-II; Udalski, Kubiak and 
Szyma{\'n}ski 1997). About 2.4 square degrees of central part of the SMC were 
observed each night during the observing seasons 1997--2000. Photometry was 
obtained with the $BVI$ filters, closely resembling the standard system. 

The OGLE-II survey has yielded a~particularly rich harvest of variable stars 
from the SMC. In the previous papers we presented the catalog of eclipsing 
binary stars (Udalski \etal 1998b), the catalog of Cepheids from the SMC 
(Udalski \etal 1999), and general catalog of variable stars detected in the 
Magellanic Clouds ({\.Z}ebru{\'n} \etal 2001). In addition {\it BVI} maps of 
the SMC were released providing precise photometry and astrometry of about 2.2 
million stars (Udalski \etal 1998a). 

In this paper we present a sample of 571 RR~Lyr stars and several other 
pulsating objects, likely $\delta$~Sct stars, detected in the OGLE-II fields 
in the SMC. The stars were selected from the reprocessed OGLE-II photometry 
based on the Difference Image Analysis (DIA) technique -- Wo{\'z}niak's (2000) 
implementation of Alard and Lupton (1998) and Alard (2000) optimal Point 
Spread Function matching algorithm. 

Similarly to the previous catalogs, all data presented in this paper, 
including individual observations, are available to the astronomical community 
from the OGLE {\sc Internet} archive. 

\Section{Observations and Data Reductions}
Observations presented in this paper were collected during the second phase of 
the OGLE microlensing search with the 1.3-m Warsaw telescope at Las Campanas 
Observatory, Chile. The observatory is operated by the Carnegie Institution of 
Washington. The telescope was equipped with the ``first generation'' camera 
with a~SITe ${2048\times2048}$ CCD detector working in drift-scan mode. The 
pixel size was 24~$\mu$m giving the 0.417~arcsec/pixel scale. Observations of 
the SMC were performed in the ``slow'' reading mode of the CCD detector with 
the gain 3.8~e$^-$/ADU and readout noise of about 5.4~e$^-$. Details of the 
instrumentation setup can be found in Udalski, Kubiak and Szyma{\'n}ski 
(1997). 

Observations of the SMC were collected between June 26, 1997 and November 25, 
2000. Eleven driftscan fields (SMC$\_$SC1--SMC$\_$SC11) covering about 
2.4~square degrees of central parts of the SMC were observed. The majority of 
frames were taken in the $I$ photometric band (about 280--340 epochs depending 
on the field). Other images were collected through the {\it V}-band (typically 
about 30 epochs) and {\it B}-band (about 20 epochs) filters. The effective 
exposure time lasted 125, 174 and~237~seconds for the $I$, $V$ and 
{\it B}-band, respectively. The median seeing was about 1\zdot\arcs3 for our 
dataset. 

The {\it I}-band photometry was obtained using Difference Images Analysis 
(DIA) -- image subtraction algorithm developed by Alard and Lupton (1998) and 
Alard (2000), and implemented by Wo{\'z}niak (2000). We introduced several 
modifications compared to the DIA techniques employed in a catalog of variable 
stars in the Magellanic Clouds ({\.Z}ebru{\'n} \etal 2001). For instance, we 
performed the DIA photometry for each star found in the reference image 
instead of the variable objects only. Further details of the DIA analysis and 
calibration of photometry may be found in the papers by Wo{\'z}niak (2000) and 
by {\.Z}ebru{\'n}, Soszy{\'n}ski and Wo{\'z}niak (2001). 

The frames in the $V$ and $B$ bands were analyzed using the {\sc DoPhot} 
photometry program (Schechter, Saha and Mateo 1993). Transformation of the 
instrumental photometry to the standard system is described by Udalski \etal 
(1998a). 

Equatorial coordinates of all stars were calculated in the identical manner as 
described in Udalski \etal (1998a). The internal accuracy of the determined 
equatorial coordinates, as measured in the overlapping regions of neighboring 
fields, is about 0\zdot\arcs15--0\zdot\arcs20 with possible systematic errors 
of the DSS coordinate system up to 0\zdot\arcs7.

\begin{figure}[htb]
\centerline{\includegraphics[width=12cm]{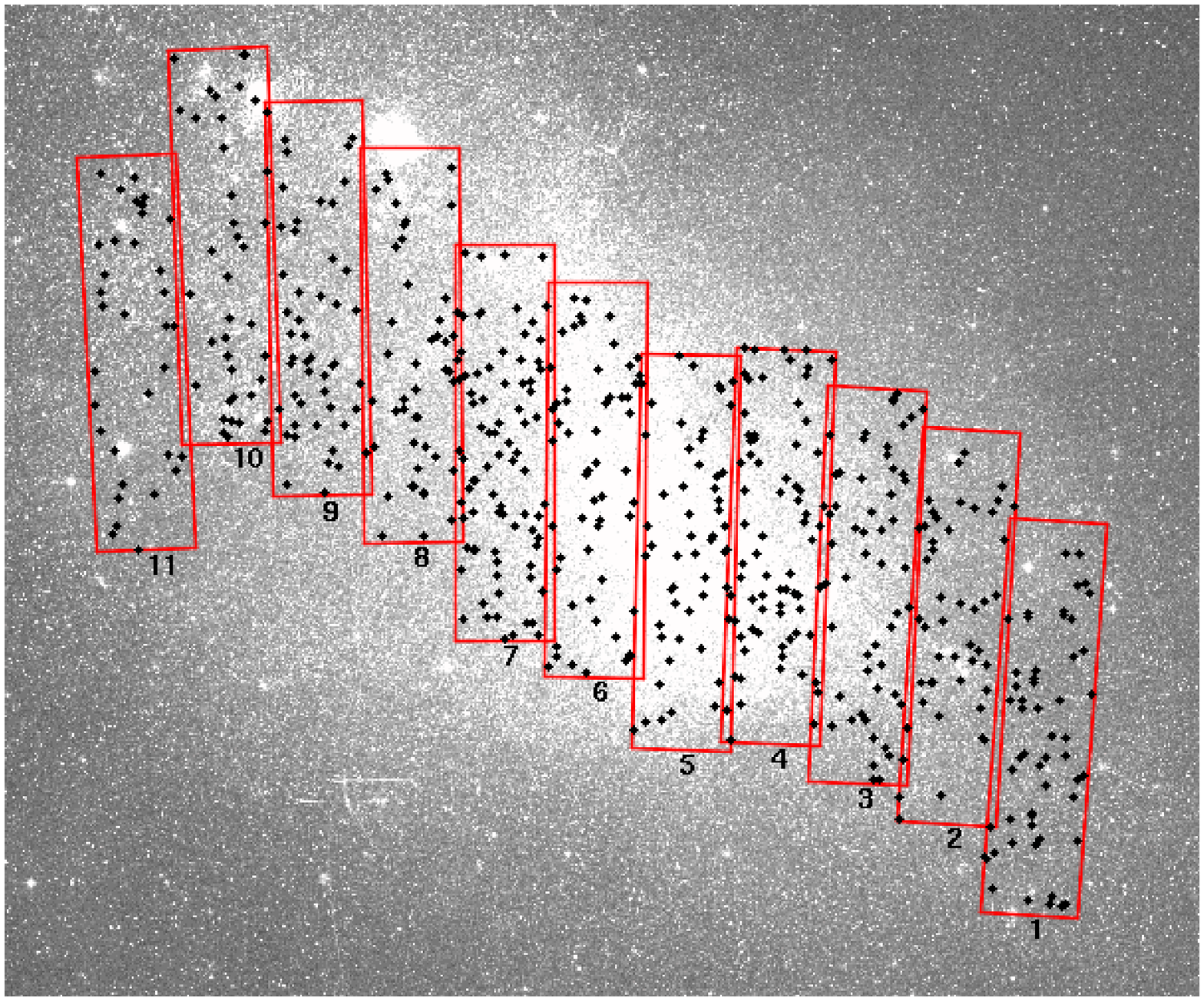}}
\vskip3mm
\FigCap{OGLE-II fields in the SMC. Dots indicate positions of RR~Lyr stars 
from the Catalog. North is up and East to the left in this Digitized Sky 
Survey image of the SMC.} 
\end{figure}
Fig.~1 presents the picture of the SMC from Digitized Sky Survey CD-ROMs with 
contours of the OGLE-II fields. Positions of RR~Lyr stars are marked with 
black dots. One can easily notice that the surface density of RR~Lyr variables 
is very uniform within the galaxy. This confirms the results of Graham (1975) 
and Smith \etal (1992) who noticed that RR~Lyr stars are not strongly 
concentrated toward either the bar or the center of the SMC. 

\Section{Interstellar Reddening}
Determination of the interstellar reddening to the SMC fields was performed by 
Udalski \etal (1999). In short, they used red clump stars for mapping the 
fluctuations of mean reddening in OGLE fields, treating their mean 
{\it I}-band magnitude as the reference brightness. Since reddening in the SMC 
is smaller and more homogeneous than in the LMC, it was determined only in 11 
lines-of-sight -- one per OGLE field. The zero point of the reddening map was 
derived on the basis of previous determinations toward two stars clusters: 
NGC416 and NGC330. 
\renewcommand{\arraystretch}{.9}
\MakeTable{lc}{12.5cm}{$E(B-V)$ reddening in the SMC fields}
{\hline
\noalign{\vskip2pt}
\multicolumn{1}{c}{Field}& $E(B-V)$\\
\noalign{\vskip1pt}
\hline
\noalign{\vskip2pt}
SMC$\_$SC1 & 0.070\\ SMC$\_$SC2 & 0.078\\ SMC$\_$SC3 & 0.089\\ SMC$\_$SC4 &
0.094\\ SMC$\_$SC5 & 0.101\\ SMC$\_$SC6 & 0.094\\ SMC$\_$SC7 & 0.097\\
SMC$\_$SC8 & 0.100\\ SMC$\_$SC9 & 0.076\\ SMC$\_$SC10& 0.079\\ SMC$\_$SC11&
0.084\\
\hline}

The final $E(B-V)$ reddening in the SMC is listed in Table~1. The error of the 
map is equal to $\pm0.02$~mag. Similar values of reddening were estimated by 
Sharpee \etal (2002) for a field in the northeast arm of the SMC located close 
to the OGLE fields SMC$\_$SC10 and SMC$\_$SC11. Interstellar extinction in the 
{\it BVI} bands can be calculated using the standard extinction curve 
coefficients (\eg Schlegel \etal 1998):
$$A_B=4.32\cdot E(B-V),\quad A_V=3.24\cdot E(B-V),\quad A_I=1.96\cdot E(B-V)$$

\vspace*{-12pt}
\Section{Selection of RR Lyr Stars}
\vspace*{-3pt}
Selection of RR~Lyr stars was performed in two stages. First, preliminary 
search for variable stars was performed using the regular OGLE-II PSF ({\sc 
DoPhot}) photometry. The mean {\it I}-band magnitude of objects was limited to 
${I<20}$~mag. The minimal number of individual measurements was set to 50. 
Candidates for variable stars were selected based on comparison of the 
standard deviation of photometry with typical standard deviation for stars 
with similar brightness. Light curves of selected candidates were searched for 
periodicity using the AoV algorithm (Schwarzenberg-Czerny 1989). Light curves 
of all objects revealing statistically significant periodic signal were then 
visually inspected. Candidates for RR~Lyr stars were extracted on the basis of 
light curve shapes and magnitudes of the stars. 

The second, final search for RR~Lyr variables was performed with the DIA 
photometry of all stars detected in the reference images. The quality of DIA 
photometry is improved by a factor of at least two, compared to the {\sc 
DoPhot} photometry (Wo{\'z}niak 2000), what enabled us to increase 
considerably the completeness of the catalog. 

We examined with the AoV period search technique the DIA photometry of objects 
with the mean magnitude ${18.4<I<19.4}$ and standard deviation at least 
0.02~mag larger than typical standard deviation for non-variable stars of 
similar brightness. Additionally we checked stars with the {\it V}-band 
magnitude between 18.9 and 20.0 and the standard deviation of {\it V}-band 
photometry 0.05~mag larger than typical. New candidates for RR~Lyr stars were 
selected from a sample of stars with periods smaller than 1 day based on 
visual inspection of their light curves and location in the color-magnitude 
diagram (CMD). 

The second stage of variability search increased the number of candidates for 
RR~Lyr stars by about 30\%. In total 571 candidates for RR~Lyr stars were 
identified. Several objects with colors outside the range ${0.2<V-I<0.8}$ and 
objects with no color information but with evident RR~Lyr-type light curves 
were also included to this sample. They can be highly reddened or blended 
stars. 

\Section{Classification}
We divided all objects into four groups: fundamental mode RR~Lyr stars (RRab), 
first overtone (RRc), double mode RR~Lyr stars (RRd) and other variable stars, 
which, in the majority of cases, are probably $\delta$~Sct stars. 

Selection of RRab-type stars from our sample was not difficult, because these 
variables form well separated groups in various diagrams. We decided to use 
period--amplitude diagram to classify fundamental-mode pulsators, because 
periods and amplitudes are the observables that are measured with the highest 
precision. 

There were more problems with separation of the first overtone RR~Lyr stars 
and short-period variable stars. Both groups of objects in our sample have 
similar luminosities, amplitudes and shapes of light curve. Additionally, the 
range of periods of $\delta$~Sct stars overlaps with the range of periods of 
RRc variables. 

The models of Bono \etal (1997) predict that the first-overtone RR~Lyr stars 
with the shortest periods should have the lowest amplitudes. Indeed, in the 
period--amplitude diagrams (Fig.~4) one can notice a minimum of {\it IV}-band 
amplitudes near the period of about 0.26~days. Therefore, we decided to mark 
as the first overtone RR~Lyr candidates with periods larger than 0.26~days. 
Objects with shorter periods were classified as $\delta$~Sct stars. However, 
we stress that it is possible that some of the RRc variables on our list are 
$\delta$~Sct or other types of variable stars, and {\it vice versa}. It is 
also possible that some second-harmonic RR~Lyr stars (RRe) can be included 
in the short period group. Further observations, especially in the Str\"omgren 
{\it uvby} filters, could provide additional information for final 
classification. 

We used two methods to search for multi-periodic variable stars. First, we 
fitted a fourth order Fourier series to each folded light curve from our 
sample and subtracted fitted function from the observational data. Then, the 
residuals were searched for periodic signal and, if detected, such a candidate 
was marked for visual inspection. 

Second search for double-mode RR~Lyr stars was performed using the {\sc Clean} 
algorithm of period determination (Roberts, Leh{\'a}r and Dreher 1987). All 
RR~Lyr candidates from our list were subject to the {\sc Clean} period 
analysis. In further analysis we selected objects for which the ratio of the 
highest peak in the power spectrum and one of the next four strongest peaks 
was close to 0.745. 

Final list of the double-mode variable stars presented in this paper was 
obtained after careful visual inspection of power spectra and folded light 
curves of candidates. We identified 57 RRd stars, 2 double mode $\delta$~Sct 
stars and several dozen stars with two closely spaced frequencies, that are 
probably members of a new class of multi-periodic RR~Lyr stars described by 
Olech \etal (1999). About 10\% in both, RRab and RRc groups, exhibited 
secondary periodicity very close to the primary pulsation frequency, with 
period ratios in the range 0.98--1.02. This frequency pattern cannot be 
explained by a superposition of radial pulsations and is therefore believed 
to be related to non-radial modes. Non-radial oscillations have also been 
detected in some RR~Lyr stars in the Galactic bulge by Moskalik (2000) and in 
the LMC by Kov{\'a}cs \etal (2000). Theoretical models of these stars have 
been proposed by Dziembowski and Cassisi (1999). 

\Section{Catalog of RR Lyr Stars}
\Subsection{Single-Mode RR Lyr Stars}
458 RRab and 56 RRc variable stars passed our selection criteria. They are 
listed in Tables~2 and~3. First two columns of both tables contain the star 
identification: field\_name star\_ID. The star\_IDs are simultaneously 
equatorial coordinates, RA and DEC (J2000), of the objects. In the next 
columns period in days and moment of the zero phase corresponding to maximum 
light are given. Finally, last columns show intensity mean {\it IVB} 
photometry. 
\renewcommand{\arraystretch}{0.8}
\renewcommand{\TableFont}{\scriptsize}
\setcounter{table}{1}
\MakeTableSepp{
l@{\hspace{4pt}}
l@{\hspace{4pt}}
c@{\hspace{4pt}}
c@{\hspace{4pt}}
c@{\hspace{4pt}}
c@{\hspace{4pt}}
c@{\hspace{4pt}}}
{12.5cm}{ab-type RR Lyrae stars from the SMC}
{\hline
\noalign{\vskip3pt}
\multicolumn{1}{c}{Field} & \multicolumn{1}{c}{Star ID} & 
$P$ & $T_0$ & $I$ & $V$ & $B$ \\
& & [days] & [HJD] & [mag] & [mag] & [mag] \\
\noalign{\vskip3pt}
\hline
\noalign{\vskip3pt}
SMC$\_$SC1  & OGLE003707.89$-$735613.9 & 0.628518 & 2450450.08631 & 18.91 & 19.46 & 19.80 \\
SMC$\_$SC1  & OGLE003642.77$-$735612.1 & 0.525673 & 2450450.01915 & 19.25 & 19.79 &  --   \\
SMC$\_$SC1  & OGLE003752.40$-$735555.9 & 0.577047 & 2450450.15096 & 18.99 & 19.60 & 19.82 \\
SMC$\_$SC1  & OGLE003634.14$-$735553.1 & 0.633026 & 2450450.33529 & 19.12 & 19.76 & 20.09 \\
SMC$\_$SC1  & OGLE003707.32$-$735455.8 & 0.562787 & 2450450.46047 & 19.17 & 19.66 & 20.00 \\
SMC$\_$SC1  & OGLE003925.15$-$735015.8 & 0.689856 & 2450450.07815 & 18.61 & 19.30 & 19.78 \\
SMC$\_$SC1  & OGLE003909.14$-$734921.4 & 0.564473 & 2450450.45890 & 18.70 & 19.23 & 19.55 \\
SMC$\_$SC1  & OGLE003836.07$-$734750.2 & 0.583830 & 2450450.12202 & 19.19 & 19.80 & 20.12 \\
SMC$\_$SC1  & OGLE003623.02$-$734651.0 & 0.565954 & 2450450.30161 & 19.01 & 19.51 & 19.83 \\
SMC$\_$SC1  & OGLE003920.43$-$734540.7 & 0.600827 & 2450450.56287 & 19.07 & 19.69 & 19.96 \\
SMC$\_$SC1  & OGLE003756.32$-$734508.8 & 0.588990 & 2450450.32308 & 19.12 & 19.75 & 20.09 \\
SMC$\_$SC1  & OGLE003841.21$-$734422.9 & 0.410569 & 2450450.35916 & 19.54 & 20.02 & 20.22 \\
SMC$\_$SC1  & OGLE003758.84$-$734339.5 & 0.523207 & 2450450.04769 & 19.11 & 19.60 & 19.93 \\
SMC$\_$SC1  & OGLE003757.71$-$734326.9 & 0.583903 & 2450450.37033 & 18.70 & 19.24 & 19.55 \\
SMC$\_$SC1  & OGLE003737.63$-$733914.6 & 0.588109 & 2450450.45983 & 19.18 & 19.70 & 20.07 \\
SMC$\_$SC1  & OGLE003633.22$-$733800.8 & 0.630791 & 2450450.32372 & 19.02 & 19.57 & 19.93 \\
SMC$\_$SC1  & OGLE003620.57$-$733724.0 & 0.639572 & 2450450.05727 & 19.12 & 19.70 & 20.20 \\
SMC$\_$SC1  & OGLE003847.37$-$733715.8 & 0.698386 & 2450450.36581 & 19.05 & 19.68 & 20.08 \\
SMC$\_$SC1  & OGLE003750.29$-$733633.9 & 0.575954 & 2450450.30285 & 18.93 & 19.46 & 19.71 \\
SMC$\_$SC1  & OGLE003831.01$-$733539.7 & 0.613462 & 2450450.36119 & 19.14 & 19.71 & 20.07 \\
SMC$\_$SC1  & OGLE003739.99$-$733513.5 & 0.667574 & 2450450.22469 & 19.11 & 19.75 & 20.14 \\
SMC$\_$SC1  & OGLE003824.91$-$733500.5 & 0.602657 & 2450450.33238 & 19.07 & 19.64 & 19.92 \\
SMC$\_$SC1  & OGLE003731.46$-$733210.8 & 0.687784 & 2450450.57565 & 18.84 & 19.44 & 19.81 \\
SMC$\_$SC1  & OGLE003834.63$-$732829.9 & 0.608249 & 2450450.33384 & 18.83 & 19.32 & 19.61 \\
SMC$\_$SC1  & OGLE003901.88$-$732727.5 & 0.582833 & 2450450.37265 & 18.73 & 19.39 & 19.32 \\
SMC$\_$SC1  & OGLE003839.07$-$732726.0 & 0.547269 & 2450450.31495 & 19.19 & 19.70 & 20.03 \\
SMC$\_$SC1  & OGLE003723.52$-$732700.6 & 0.522510 & 2450450.02864 & 19.21 & 19.70 & 20.07 \\
SMC$\_$SC1  & OGLE003851.89$-$732530.1 & 0.593166 & 2450450.04659 & 19.18 & 19.68 & 19.99 \\
SMC$\_$SC1  & OGLE003816.46$-$732449.2 & 0.427897 & 2450450.16153 & 19.45 & 19.77 & 20.04 \\
SMC$\_$SC1  & OGLE003815.86$-$732403.7 & 0.588187 & 2450450.29876 & 19.16 & 19.97 & 20.00 \\
SMC$\_$SC1  & OGLE003814.29$-$732257.8 & 0.485913 & 2450450.12678 & 19.14 & 19.63 & 19.94 \\
SMC$\_$SC1  & OGLE003700.85$-$732038.6 & 0.496091 & 2450450.20237 & 19.39 & 19.90 & 20.17 \\
SMC$\_$SC1  & OGLE003640.43$-$731943.1 & 0.642239 & 2450450.15996 & 18.97 & 19.52 & 19.90 \\
SMC$\_$SC1  & OGLE003915.76$-$731555.1 & 0.517790 & 2450450.04688 & 19.13 & 19.64 & 19.91 \\
SMC$\_$SC1  & OGLE003836.32$-$731526.8 & 0.636549 & 2450450.25708 & 19.13 & 19.82 & 20.18 \\
SMC$\_$SC1  & OGLE003711.49$-$731513.2 & 0.657652 & 2450450.45133 & 18.99 & 19.63 &  --   \\
SMC$\_$SC1  & OGLE003727.78$-$731454.9 & 0.412698 & 2450450.33535 & 19.51 & 20.02 & 20.29 \\
SMC$\_$SC1  & OGLE003726.21$-$731420.0 & 0.519471 & 2450450.31104 & 19.13 & 19.59 & 19.90 \\
SMC$\_$SC1  & OGLE003641.77$-$731120.1 & 0.598166 & 2450450.51871 & 19.04 & 19.57 &  --   \\
SMC$\_$SC1  & OGLE003704.40$-$731023.2 & 0.651941 & 2450450.55886 & 18.78 & 19.31 & 19.72 \\
SMC$\_$SC1  & OGLE003648.60$-$731003.0 & 0.554478 & 2450450.06475 & 19.30 & 19.95 &  --   \\
SMC$\_$SC2  & OGLE004231.36$-$734227.2 & 0.551124 & 2450450.31254 & 19.13 & 19.72 &  --   \\
SMC$\_$SC2  & OGLE004227.44$-$733655.1 & 0.658287 & 2450450.65074 & 19.03 & 19.64 & 20.04 \\
SMC$\_$SC2  & OGLE003951.38$-$733307.7 & 0.520450 & 2450450.20499 & 19.16 & 19.64 & 19.89 \\
SMC$\_$SC2  & OGLE004123.87$-$733024.5 & 0.633188 & 2450450.06167 & 18.78 & 19.35 &  --   \\
SMC$\_$SC2  & OGLE003959.69$-$732855.5 & 0.639105 & 2450450.07035 & 19.20 & 19.81 & 20.23 \\
SMC$\_$SC2  & OGLE004156.67$-$732831.6 & 0.568383 & 2450450.03657 & 18.90 & 19.43 & 19.74 \\
SMC$\_$SC2  & OGLE003950.33$-$732627.0 & 0.635539 & 2450450.13159 & 19.08 & 19.72 & 20.11 \\
SMC$\_$SC2  & OGLE004143.00$-$732553.5 & 0.570230 & 2450450.51291 & 19.05 & 19.48 & 19.83 \\
SMC$\_$SC2  & OGLE004138.10$-$732540.6 & 0.555241 & 2450450.28273 & 19.27 & 19.79 & 20.10 \\
SMC$\_$SC2  & OGLE004043.07$-$732509.1 & 0.649512 & 2450450.21117 & 19.00 & 19.62 & 20.17 \\
SMC$\_$SC2  & OGLE003953.93$-$732246.5 & 0.604904 & 2450450.04027 & 19.01 & 19.53 & 19.90 \\
SMC$\_$SC2  & OGLE004022.62$-$732222.4 & 0.580435 & 2450450.26546 & 19.23 & 19.79 & 20.13 \\
SMC$\_$SC2  & OGLE004129.39$-$732202.3 & 0.629913 & 2450450.18934 & 19.24 & 19.86 & 20.04 \\
SMC$\_$SC2  & OGLE004059.14$-$732155.8 & 0.749157 & 2450450.38856 & 19.15 & 19.76 & 20.23 \\
SMC$\_$SC2  & OGLE004010.03$-$732044.6 & 0.583876 & 2450450.38355 & 19.44 & 20.05 & 20.38 \\
SMC$\_$SC2  & OGLE004027.63$-$731954.9 & 0.595305 & 2450450.26668 & 19.44 & 20.02 & 20.44 \\
SMC$\_$SC2  & OGLE004202.17$-$731758.8 & 0.572903 & 2450450.15389 & 19.04 & 19.60 & 20.00 \\
SMC$\_$SC2  & OGLE004025.49$-$731452.5 & 0.605694 & 2450450.58069 & 19.25 & 19.83 & 20.28 \\
SMC$\_$SC2  & OGLE004131.78$-$731328.5 & 0.604973 & 2450450.23258 & 19.02 & 19.59 &  --   \\
SMC$\_$SC2  & OGLE003943.19$-$731245.9 & 0.593981 & 2450450.28635 & 18.91 & 19.32 & 19.56 \\
SMC$\_$SC2  & OGLE004204.03$-$730813.5 & 0.636453 & 2450450.29674 & 19.17 & 19.75 & 20.24 \\
SMC$\_$SC2  & OGLE004156.46$-$730641.7 & 0.586298 & 2450450.17198 & 19.13 &  --   &  --   \\
SMC$\_$SC2  & OGLE004154.36$-$730539.3 & 0.524564 & 2450450.23252 & 19.27 & 19.80 & 20.18 \\
SMC$\_$SC2  & OGLE004126.84$-$730355.2 & 0.705041 & 2450450.68756 & 18.89 & 19.58 & 20.11 \\
SMC$\_$SC2  & OGLE004225.29$-$730349.8 & 0.699868 & 2450450.24522 & 18.88 & 19.34 & 19.77 \\
SMC$\_$SC2  & OGLE004211.68$-$730329.2 & 0.632047 & 2450450.21854 & 19.34 & 20.04 & 20.54 \\
SMC$\_$SC2  & OGLE004150.76$-$730220.5 & 0.578257 & 2450450.49694 & 19.19 & 19.88 & 20.34 \\
\noalign{\vskip3pt}
\hline}

\setcounter{table}{1}
\MakeTableSepp{
l@{\hspace{4pt}}
l@{\hspace{4pt}}
c@{\hspace{4pt}}
c@{\hspace{4pt}}
c@{\hspace{4pt}}
c@{\hspace{4pt}}
c@{\hspace{4pt}}}
{12.5cm}{Continued}
{\hline
\noalign{\vskip3pt}
\multicolumn{1}{c}{Field} & \multicolumn{1}{c}{Star ID} & 
$P$ & $T_0$ & $I$ & $V$ & $B$ \\
& & [days] & [HJD] & [mag] & [mag] & [mag] \\
\noalign{\vskip3pt}
\hline
\noalign{\vskip3pt}
SMC$\_$SC2  & OGLE003922.76$-$725950.6 & 0.544786 & 2450450.13952 & 19.21 & 19.85 &  --   \\
SMC$\_$SC2  & OGLE004206.40$-$725949.0 & 0.608147 & 2450450.04027 & 19.19 & 19.86 & 20.29 \\
SMC$\_$SC2  & OGLE004106.44$-$725948.4 & 0.510413 & 2450450.35386 & 19.28 & 19.83 & 20.21 \\
SMC$\_$SC2  & OGLE003952.24$-$725901.7 & 0.591274 & 2450450.33622 & 19.12 & 19.67 & 19.99 \\
SMC$\_$SC2  & OGLE003947.05$-$725700.1 & 0.554868 & 2450450.28534 & 19.06 & 19.56 & 19.90 \\
SMC$\_$SC2  & OGLE004105.74$-$725238.1 & 0.539650 & 2450450.52923 & 18.91 & 19.38 & 19.66 \\
SMC$\_$SC3  & OGLE004324.97$-$734011.6 & 0.585206 & 2450450.01341 & 19.08 &  --   & 20.01 \\
SMC$\_$SC3  & OGLE004312.60$-$734000.8 & 0.508109 & 2450450.48203 & 19.24 & 19.86 & 20.08 \\
SMC$\_$SC3  & OGLE004326.58$-$733809.2 & 0.583202 & 2450450.41598 & 19.20 & 19.80 & 20.36 \\
SMC$\_$SC3  & OGLE004227.44$-$733655.1 & 0.658282 & 2450450.00702 & 19.05 & 19.64 & 20.05 \\
SMC$\_$SC3  & OGLE004258.04$-$733642.7 & 0.636162 & 2450450.18009 & 18.91 & 19.41 & 19.83 \\
SMC$\_$SC3  & OGLE004306.71$-$733527.9 & 0.466095 & 2450450.33967 & 19.60 & 19.98 & 20.32 \\
SMC$\_$SC3  & OGLE004454.78$-$733241.3 & 0.567085 & 2450450.02131 & 19.14 & 19.72 & 20.08 \\
SMC$\_$SC3  & OGLE004533.91$-$733240.3 & 0.594071 & 2450450.12923 & 19.37 & 19.87 & 20.16 \\
SMC$\_$SC3  & OGLE004416.71$-$733148.7 & 0.674706 & 2450450.14972 & 19.24 & 19.94 & 20.40 \\
SMC$\_$SC3  & OGLE004350.56$-$733141.1 & 0.588695 & 2450450.18045 & 19.06 & 19.66 & 20.03 \\
SMC$\_$SC3  & OGLE004356.46$-$733045.4 & 0.535740 & 2450450.51319 & 19.28 & 19.88 & 20.30 \\
SMC$\_$SC3  & OGLE004238.25$-$732911.4 & 0.578983 & 2450450.18109 & 19.28 & 19.90 & 20.39 \\
SMC$\_$SC3  & OGLE004526.76$-$732801.7 & 0.613725 & 2450450.55600 & 19.20 & 19.92 & 20.31 \\
SMC$\_$SC3  & OGLE004329.74$-$732705.8 & 0.564446 & 2450450.42257 & 19.20 & 19.72 & 20.13 \\
SMC$\_$SC3  & OGLE004531.73$-$732634.2 & 0.507230 & 2450450.35151 & 19.41 & 19.92 & 20.23 \\
SMC$\_$SC3  & OGLE004323.58$-$732338.5 & 0.631343 & 2450450.27193 & 19.06 & 19.63 & 20.00 \\
SMC$\_$SC3  & OGLE004351.26$-$732225.9 & 0.652652 & 2450450.15028 & 19.33 & 20.05 &  --   \\
SMC$\_$SC3  & OGLE004506.65$-$731845.6 & 0.667459 & 2450450.17054 & 19.28 & 20.12 & 20.59 \\
SMC$\_$SC3  & OGLE004307.06$-$731808.6 & 0.581089 & 2450450.30603 & 19.14 & 19.65 & 19.80 \\
SMC$\_$SC3  & OGLE004256.85$-$731612.8 & 0.516150 & 2450450.04373 & 19.59 & 20.16 & 20.55 \\
SMC$\_$SC3  & OGLE004236.24$-$731512.1 & 0.593325 & 2450450.16000 & 19.15 & 19.83 & 20.35 \\
SMC$\_$SC3  & OGLE004533.00$-$731255.1 & 0.601841 & 2450450.23558 & 18.86 & 19.38 &  --   \\
SMC$\_$SC3  & OGLE004520.13$-$731233.5 & 0.579379 & 2450450.55955 & 19.52 & 20.32 & 20.75 \\
SMC$\_$SC3  & OGLE004441.67$-$731126.2 & 0.760147 & 2450450.08199 & 18.94 & 19.57 & 19.98 \\
SMC$\_$SC3  & OGLE004336.02$-$730919.6 & 0.613051 & 2450450.34229 & 19.08 & 19.66 & 19.99 \\
SMC$\_$SC3  & OGLE004405.30$-$730800.8 & 0.559465 & 2450450.00829 & 19.54 & 19.98 & 20.12 \\
SMC$\_$SC3  & OGLE004514.95$-$730638.7 & 0.619373 & 2450450.53815 & 18.78 & 19.55 &  --   \\
SMC$\_$SC3  & OGLE004414.01$-$730449.8 & 0.474392 & 2450450.19235 & 19.36 & 19.89 & 20.32 \\
SMC$\_$SC3  & OGLE004339.08$-$730426.5 & 0.588119 & 2450450.05648 & 19.17 & 19.69 & 20.05 \\
SMC$\_$SC3  & OGLE004429.26$-$730421.4 & 0.520797 & 2450450.22782 & 18.69 & 19.10 & 19.53 \\
SMC$\_$SC3  & OGLE004225.29$-$730349.8 & 0.699864 & 2450450.26169 & 18.82 &  --   &  --   \\
SMC$\_$SC3  & OGLE004315.39$-$730300.3 & 0.624452 & 2450450.40185 & 19.09 & 19.65 & 20.15 \\
SMC$\_$SC3  & OGLE004520.12$-$730224.8 & 0.659571 & 2450450.59086 & 19.26 & 20.03 & 20.52 \\
SMC$\_$SC3  & OGLE004508.81$-$730126.7 & 0.700450 & 2450450.61532 & 19.02 & 19.49 & 19.76 \\
SMC$\_$SC3  & OGLE004533.33$-$730120.2 & 0.572240 & 2450450.37587 & 18.99 &  --   &  --   \\
SMC$\_$SC3  & OGLE004422.52$-$725710.3 & 0.656097 & 2450450.60125 & 19.07 & 19.69 & 20.08 \\
SMC$\_$SC3  & OGLE004255.30$-$725707.7 & 0.590113 & 2450450.31280 & 19.18 & 19.74 & 20.09 \\
SMC$\_$SC3  & OGLE004512.96$-$725645.9 & 0.552929 & 2450450.21355 & 19.29 & 19.83 & 20.23 \\
SMC$\_$SC3  & OGLE004300.35$-$725626.0 & 0.543195 & 2450450.14120 & 18.91 & 19.36 & 19.64 \\
SMC$\_$SC3  & OGLE004504.31$-$725623.9 & 0.567931 & 2450450.21802 & 19.08 & 19.62 & 19.92 \\
SMC$\_$SC3  & OGLE004319.78$-$725528.9 & 0.589635 & 2450450.04537 & 19.04 & 19.57 & 19.95 \\
SMC$\_$SC3  & OGLE004333.99$-$725339.4 & 0.616317 & 2450450.37327 & 19.19 & 19.82 & 20.20 \\
SMC$\_$SC3  & OGLE004335.00$-$725231.3 & 0.586273 & 2450450.07666 & 19.27 & 19.80 & 20.20 \\
SMC$\_$SC3  & OGLE004422.31$-$725127.9 & 0.645223 & 2450450.38919 & 19.10 & 19.77 & 20.24 \\
SMC$\_$SC3  & OGLE004528.73$-$725109.7 & 0.656902 & 2450450.10323 & 19.06 & 19.71 & 20.15 \\
SMC$\_$SC3  & OGLE004420.93$-$725034.4 & 0.616291 & 2450450.60864 & 19.18 & 19.81 & 20.23 \\
SMC$\_$SC3  & OGLE004307.78$-$724922.1 & 0.585893 & 2450450.15964 & 19.20 & 19.74 & 20.17 \\
SMC$\_$SC3  & OGLE004326.82$-$724559.3 & 0.567412 & 2450450.16543 & 19.22 & 19.77 & 20.25 \\
SMC$\_$SC3  & OGLE004322.93$-$724447.5 & 0.648591 & 2450450.31631 & 18.86 & 19.45 & 19.84 \\
SMC$\_$SC4  & OGLE004819.70$-$733521.5 & 0.596133 & 2450450.46771 & 19.28 &  --   &  --   \\
SMC$\_$SC4  & OGLE004533.91$-$733240.3 & 0.594074 & 2450450.12299 & 19.35 & 19.75 & 19.89 \\
SMC$\_$SC4  & OGLE004827.91$-$733108.7 & 0.637821 & 2450450.38748 & 18.98 & 19.57 & 19.99 \\
SMC$\_$SC4  & OGLE004801.59$-$733021.5 & 0.399572 & 2450450.14594 & 19.49 & 19.91 & 20.21 \\
SMC$\_$SC4  & OGLE004526.76$-$732801.7 & 0.613714 & 2450450.56070 & 19.23 & 19.87 &  --   \\
SMC$\_$SC4  & OGLE004531.73$-$732634.2 & 0.507225 & 2450450.37273 & 19.46 & 19.88 & 20.30 \\
SMC$\_$SC4  & OGLE004802.63$-$732627.0 & 0.610070 & 2450450.29492 & 18.96 & 19.56 & 19.92 \\
SMC$\_$SC4  & OGLE004816.17$-$732618.9 & 0.599649 & 2450450.57126 & 19.29 & 20.14 & 20.38 \\
SMC$\_$SC4  & OGLE004601.00$-$732535.3 & 0.588723 & 2450450.17922 & 18.96 & 19.50 & 19.82 \\
SMC$\_$SC4  & OGLE004644.13$-$732500.7 & 0.622674 & 2450450.45496 & 18.25 & 18.42 & 18.51 \\
SMC$\_$SC4  & OGLE004642.76$-$732311.2 & 0.619927 & 2450450.18659 & 18.97 & 19.80 & 20.16 \\
SMC$\_$SC4  & OGLE004758.98$-$732241.3 & 0.449794 & 2450450.18042 & 19.80 & 20.46 & 20.91 \\
SMC$\_$SC4  & OGLE004644.71$-$732140.8 & 0.574656 & 2450450.46645 & 19.11 & 19.51 & 19.70 \\
\noalign{\vskip3pt}
\hline}

\setcounter{table}{1}
\MakeTableSepp{
l@{\hspace{4pt}}
l@{\hspace{4pt}}
c@{\hspace{4pt}}
c@{\hspace{4pt}}
c@{\hspace{4pt}}
c@{\hspace{4pt}}
c@{\hspace{4pt}}}
{12.5cm}{Continued}
{\hline
\noalign{\vskip3pt}
\multicolumn{1}{c}{Field} & \multicolumn{1}{c}{Star ID} & 
$P$ & $T_0$ & $I$ & $V$ & $B$ \\
& & [days] & [HJD] & [mag] & [mag] & [mag] \\
\noalign{\vskip3pt}
\hline
\noalign{\vskip3pt}
SMC$\_$SC4  & OGLE004626.35$-$732039.2 & 0.613248 & 2450450.25379 & 19.39 & 20.18 & 20.74 \\
SMC$\_$SC4  & OGLE004805.65$-$732033.9 & 0.523798 & 2450450.36642 & 19.24 & 19.63 & 19.93 \\
SMC$\_$SC4  & OGLE004733.11$-$732004.8 & 0.592249 & 2450450.04479 & 19.30 & 20.02 & 20.55 \\
SMC$\_$SC4  & OGLE004733.05$-$731955.4 & 0.656125 & 2450450.57332 & 19.29 & 19.92 & 20.16 \\
SMC$\_$SC4  & OGLE004548.41$-$731955.3 & 0.695294 & 2450450.29419 & 19.36 & 19.51 & 19.98 \\
SMC$\_$SC4  & OGLE004744.69$-$731919.1 & 0.598960 & 2450450.46312 & 19.12 & 19.71 & 20.18 \\
SMC$\_$SC4  & OGLE004817.88$-$731815.2 & 0.558981 & 2450450.22375 & 19.03 & 19.61 & 19.91 \\
SMC$\_$SC4  & OGLE004833.02$-$731700.0 & 0.535606 & 2450450.11273 & 19.39 & 20.00 & 20.48 \\
SMC$\_$SC4  & OGLE004650.36$-$731652.2 & 0.630375 & 2450450.32362 & 19.00 & 19.60 & 20.00 \\
SMC$\_$SC4  & OGLE004726.08$-$731632.0 & 0.563371 & 2450450.25654 & 19.05 & 20.02 &  --   \\
SMC$\_$SC4  & OGLE004611.69$-$731625.0 & 0.554438 & 2450450.33012 & 19.39 & 20.06 & 20.55 \\
SMC$\_$SC4  & OGLE004649.00$-$731544.5 & 0.665961 & 2450450.59371 & 18.98 & 19.59 & 19.89 \\
SMC$\_$SC4  & OGLE004828.10$-$731442.2 & 0.586266 & 2450450.58390 & 18.74 & 19.28 & 19.59 \\
SMC$\_$SC4  & OGLE004724.96$-$731427.2 & 0.614266 & 2450450.06254 & 18.95 & 19.54 & 19.95 \\
SMC$\_$SC4  & OGLE004623.27$-$731356.1 & 0.584151 & 2450450.53290 & 19.32 & 19.89 & 20.18 \\
SMC$\_$SC4  & OGLE004651.52$-$731333.7 & 0.487166 & 2450450.27109 & 19.79 & 20.36 & 20.79 \\
SMC$\_$SC4  & OGLE004835.73$-$731330.1 & 0.584923 & 2450450.34526 & 19.21 & 19.96 &  --   \\
SMC$\_$SC4  & OGLE004639.18$-$731324.7 & 0.424320 & 2450450.06714 & 19.60 & 20.04 & 20.34 \\
SMC$\_$SC4  & OGLE004533.00$-$731255.1 & 0.601843 & 2450450.23359 & 18.88 & 19.41 & 19.80 \\
SMC$\_$SC4  & OGLE004809.48$-$731209.0 & 0.606119 & 2450450.18750 & 19.39 & 19.78 & 20.21 \\
SMC$\_$SC4  & OGLE004721.26$-$731135.5 & 0.462014 & 2450450.10985 & 19.46 & 19.97 & 20.84 \\
SMC$\_$SC4  & OGLE004632.60$-$731109.4 & 0.685350 & 2450450.05829 & 19.02 & 19.64 & 20.04 \\
SMC$\_$SC4  & OGLE004739.98$-$730843.0 & 0.590461 & 2450450.09431 & 18.89 & 19.29 & 19.63 \\
SMC$\_$SC4  & OGLE004551.52$-$730620.0 & 0.618171 & 2450450.42983 & 19.07 & 19.67 & 20.14 \\
SMC$\_$SC4  & OGLE004732.54$-$730610.3 & 0.471967 & 2450450.16210 & 19.73 & 20.56 &  --   \\
SMC$\_$SC4  & OGLE004805.01$-$730426.6 & 0.545770 & 2450450.20954 & 19.33 & 19.88 & 20.27 \\
SMC$\_$SC4  & OGLE004539.62$-$730422.2 & 0.605842 & 2450450.32578 & 19.32 & 19.86 & 20.66 \\
SMC$\_$SC4  & OGLE004610.66$-$730355.6 & 0.611444 & 2450450.55907 & 19.23 &  --   &  --   \\
SMC$\_$SC4  & OGLE004607.56$-$730236.0 & 0.657592 & 2450450.14296 & 18.86 & 19.42 & 19.73 \\
SMC$\_$SC4  & OGLE004533.33$-$730120.2 & 0.572237 & 2450450.38078 & 19.04 & 19.50 & 19.83 \\
SMC$\_$SC4  & OGLE004807.67$-$725937.8 & 0.626137 & 2450450.56625 & 18.37 & 19.10 & 19.59 \\
SMC$\_$SC4  & OGLE004659.82$-$725937.7 & 0.651938 & 2450450.53632 & 18.95 & 19.82 & 19.67 \\
SMC$\_$SC4  & OGLE004619.25$-$725914.3 & 0.587185 & 2450450.46039 & 19.25 & 19.82 & 20.30 \\
SMC$\_$SC4  & OGLE004629.89$-$725841.9 & 0.563436 & 2450450.47365 & 19.44 & 20.14 & 20.56 \\
SMC$\_$SC4  & OGLE004649.36$-$725742.5 & 0.635306 & 2450450.58726 & 18.95 & 19.47 & 19.78 \\
SMC$\_$SC4  & OGLE004817.37$-$725536.5 & 0.556467 & 2450450.12401 & 19.09 & 19.97 & 20.11 \\
SMC$\_$SC4  & OGLE004758.51$-$725430.6 & 0.618294 & 2450450.39021 & 19.25 & 19.93 & 20.45 \\
SMC$\_$SC4  & OGLE004550.82$-$725342.1 & 0.612531 & 2450450.33796 & 19.08 & 19.65 & 20.06 \\
SMC$\_$SC4  & OGLE004800.43$-$725222.7 & 0.518919 & 2450450.51520 & 19.23 & 19.86 & 20.27 \\
SMC$\_$SC4  & OGLE004756.26$-$725217.6 & 0.592250 & 2450450.17243 & 19.09 & 19.70 & 20.14 \\
SMC$\_$SC4  & OGLE004754.45$-$725212.1 & 0.586127 & 2450450.34334 & 19.10 & 20.02 & 20.11 \\
SMC$\_$SC4  & OGLE004805.17$-$725144.4 & 0.509540 & 2450450.41483 & 19.24 & 19.61 & 20.04 \\
SMC$\_$SC4  & OGLE004755.07$-$725141.5 & 0.504549 & 2450450.23360 & 18.96 & 19.42 & 19.73 \\
SMC$\_$SC4  & OGLE004528.73$-$725109.7 & 0.656889 & 2450450.12283 & 19.09 & 19.70 &  --   \\
SMC$\_$SC4  & OGLE004616.38$-$724926.7 & 0.648315 & 2450450.36557 & 19.14 & 19.78 & 20.31 \\
SMC$\_$SC4  & OGLE004814.17$-$724837.4 & 0.536911 & 2450450.13189 & 19.30 & 19.91 & 20.27 \\
SMC$\_$SC4  & OGLE004808.76$-$724351.2 & 0.618879 & 2450450.32033 & 19.02 & 19.62 & 20.13 \\
SMC$\_$SC4  & OGLE004742.57$-$724315.5 & 0.552044 & 2450450.11251 & 19.32 & 19.91 & 20.21 \\
SMC$\_$SC4  & OGLE004816.18$-$724249.1 & 0.642870 & 2450450.50173 & 18.85 & 19.57 & 19.84 \\
SMC$\_$SC4  & OGLE004620.66$-$724246.6 & 0.596697 & 2450450.44950 & 18.86 & 19.42 &  --   \\
SMC$\_$SC4  & OGLE004615.10$-$724135.5 & 0.647222 & 2450450.36154 & 19.19 & 19.77 & 20.32 \\
SMC$\_$SC4  & OGLE004651.47$-$724040.8 & 0.649493 & 2450450.63333 & 18.92 & 19.51 & 19.87 \\
SMC$\_$SC4  & OGLE004532.76$-$724018.2 & 0.579954 & 2450450.00568 & 19.06 & 19.60 & 20.00 \\
SMC$\_$SC4  & OGLE004758.16$-$723929.7 & 0.549696 & 2450450.49921 & 19.04 & 19.60 &  --   \\
SMC$\_$SC4  & OGLE004621.06$-$723918.3 & 0.570390 & 2450450.54191 & 19.30 & 19.85 &  --   \\
SMC$\_$SC4  & OGLE004703.95$-$723915.8 & 0.554071 & 2450450.30318 & 18.99 & 19.49 &  --   \\
SMC$\_$SC5  & OGLE005135.39$-$733417.0 & 0.559560 & 2450450.27158 & 19.36 & 20.00 &  --   \\
SMC$\_$SC5  & OGLE005112.92$-$733302.6 & 0.586115 & 2450450.38079 & 18.43 & 19.04 & 19.41 \\
SMC$\_$SC5  & OGLE005040.93$-$733245.8 & 0.687076 & 2450450.24528 & 18.71 & 19.30 & 19.66 \\
SMC$\_$SC5  & OGLE005019.77$-$733131.4 & 0.511557 & 2450450.37167 & 18.58 & 19.29 & 19.80 \\
SMC$\_$SC5  & OGLE004827.91$-$733108.7 & 0.637818 & 2450450.37974 & 19.04 & 19.62 & 20.01 \\
SMC$\_$SC5  & OGLE004852.08$-$733034.5 & 0.563893 & 2450450.24775 & 19.26 & 19.77 & 20.27 \\
SMC$\_$SC5  & OGLE005045.05$-$732738.8 & 0.570909 & 2450450.04667 & 18.97 & 19.41 & 19.71 \\
SMC$\_$SC5  & OGLE005026.32$-$732418.2 & 0.422681 & 2450450.12730 & 19.59 & 20.11 & 20.23 \\
SMC$\_$SC5  & OGLE004924.05$-$732231.3 & 0.585875 & 2450450.18674 & 19.73 & 20.58 & 21.24 \\
SMC$\_$SC5  & OGLE005009.50$-$732105.8 & 0.599056 & 2450450.03930 & 19.26 & 20.01 & 20.59 \\
SMC$\_$SC5  & OGLE005042.77$-$732103.0 & 0.656440 & 2450450.61300 & 19.03 & 19.67 & 20.13 \\
SMC$\_$SC5  & OGLE005049.36$-$731917.6 & 0.630189 & 2450450.21295 & 19.35 & 20.05 & 20.57 \\
\noalign{\vskip3pt}
\hline}

\setcounter{table}{1}
\MakeTableSepp{
l@{\hspace{4pt}}
l@{\hspace{4pt}}
c@{\hspace{4pt}}
c@{\hspace{4pt}}
c@{\hspace{4pt}}
c@{\hspace{4pt}}
c@{\hspace{4pt}}}
{12.5cm}{Continued}
{\hline
\noalign{\vskip3pt}
\multicolumn{1}{c}{Field} & \multicolumn{1}{c}{Star ID} & 
$P$ & $T_0$ & $I$ & $V$ & $B$ \\
& & [days] & [HJD] & [mag] & [mag] & [mag] \\
\noalign{\vskip3pt}
\hline
\noalign{\vskip3pt}
SMC$\_$SC5  & OGLE004833.02$-$731800.0 & 0.535611 & 2450450.10067 & 19.38 & 19.97 & 20.36 \\
SMC$\_$SC5  & OGLE005019.26$-$731610.0 & 0.608400 & 2450450.29679 & 19.24 & 19.88 & 20.31 \\
SMC$\_$SC5  & OGLE004924.31$-$731447.2 & 0.545795 & 2450450.36610 & 19.49 & 20.14 & 20.65 \\
SMC$\_$SC5  & OGLE004828.10$-$731442.2 & 0.586267 & 2450450.58290 & 18.78 & 19.28 &  --   \\
SMC$\_$SC5  & OGLE005025.83$-$731403.0 & 0.599855 & 2450450.28478 & 19.45 & 20.17 &  --   \\
SMC$\_$SC5  & OGLE005024.07$-$731349.0 & 0.566477 & 2450450.33383 & 19.06 & 19.59 & 20.00 \\
SMC$\_$SC5  & OGLE004835.73$-$731330.1 & 0.584920 & 2450450.35112 & 19.25 & 19.84 & 20.28 \\
SMC$\_$SC5  & OGLE004922.63$-$731220.9 & 0.578897 & 2450450.00291 & 19.06 & 19.87 &  --   \\
SMC$\_$SC5  & OGLE005120.92$-$730920.1 & 0.549932 & 2450450.27920 & 19.28 & 19.96 & 20.30 \\
SMC$\_$SC5  & OGLE004940.50$-$730901.2 & 0.596414 & 2450450.11857 & 19.50 & 20.43 & 20.95 \\
SMC$\_$SC5  & OGLE004838.59$-$730820.1 & 0.648218 & 2450450.47363 & 18.84 & 19.61 & 19.93 \\
SMC$\_$SC5  & OGLE005110.48$-$730750.0 & 0.431720 & 2450450.09690 & 19.61 & 20.13 & 20.25 \\
SMC$\_$SC5  & OGLE004913.17$-$730651.0 & 0.620062 & 2450450.19862 & 19.22 & 19.91 & 20.30 \\
SMC$\_$SC5  & OGLE004951.17$-$730620.3 & 0.624850 & 2450450.48463 & 19.12 & 19.86 & 20.06 \\
SMC$\_$SC5  & OGLE004907.44$-$730617.5 & 0.625806 & 2450450.03943 & 19.57 & 20.48 & 21.06 \\
SMC$\_$SC5  & OGLE004845.03$-$730445.3 & 0.569002 & 2450450.49550 & 19.81 & 20.72 & 21.27 \\
SMC$\_$SC5  & OGLE004954.79$-$730321.3 & 0.803567 & 2450450.43899 & 19.01 & 19.66 & 20.03 \\
SMC$\_$SC5  & OGLE004859.12$-$730226.6 & 0.564530 & 2450450.16304 & 19.34 &  --   & 19.97 \\
SMC$\_$SC5  & OGLE004859.75$-$730111.2 & 0.636837 & 2450450.26670 & 19.60 & 20.43 & 20.99 \\
SMC$\_$SC5  & OGLE004906.31$-$725930.6 & 0.643771 & 2450450.14844 & 19.05 & 19.68 & 20.07 \\
SMC$\_$SC5  & OGLE004903.38$-$725750.1 & 0.606646 & 2450450.58766 & 19.14 & 19.97 & 20.05 \\
SMC$\_$SC5  & OGLE004854.14$-$725710.6 & 0.603874 & 2450450.25172 & 19.09 & 19.59 & 20.03 \\
SMC$\_$SC5  & OGLE004938.26$-$725616.5 & 0.516923 & 2450450.50677 & 19.55 & 20.16 &  --   \\
SMC$\_$SC5  & OGLE005001.12$-$725402.6 & 0.619444 & 2450450.27737 & 19.42 & 20.15 & 20.61 \\
SMC$\_$SC5  & OGLE004936.31$-$725229.2 & 0.642318 & 2450450.43365 & 19.10 & 19.72 & 20.08 \\
SMC$\_$SC5  & OGLE005126.54$-$725205.3 & 0.543700 & 2450450.26863 & 19.21 & 19.83 & 20.20 \\
SMC$\_$SC5  & OGLE004935.53$-$725113.9 & 0.634333 & 2450450.46663 & 19.11 & 19.80 & 20.21 \\
SMC$\_$SC5  & OGLE005015.78$-$725020.6 & 0.684064 & 2450450.33727 & 18.81 & 19.34 & 19.77 \\
SMC$\_$SC5  & OGLE004841.02$-$724731.3 & 0.517022 & 2450450.01453 & 19.25 & 19.81 & 20.22 \\
SMC$\_$SC5  & OGLE004907.86$-$724443.6 & 0.627245 & 2450450.18015 & 19.33 & 19.92 & 20.34 \\
SMC$\_$SC5  & OGLE004955.56$-$724200.4 & 0.662868 & 2450450.47838 & 18.88 & 19.49 & 19.91 \\
SMC$\_$SC6  & OGLE005314.86$-$732622.7 & 0.591032 & 2450450.36230 & 19.11 &  --   &  --   \\
SMC$\_$SC6  & OGLE005427.69$-$732534.0 & 0.592581 & 2450450.09052 & 19.02 & 19.64 & 20.17 \\
SMC$\_$SC6  & OGLE005415.18$-$732401.2 & 0.568968 & 2450450.15960 & 19.08 & 19.68 & 20.13 \\
SMC$\_$SC6  & OGLE005146.64$-$732350.2 & 0.589878 & 2450450.27402 & 19.17 & 19.70 & 20.22 \\
SMC$\_$SC6  & OGLE005147.81$-$732338.6 & 0.643648 & 2450450.04518 & 19.00 & 19.62 & 20.10 \\
SMC$\_$SC6  & OGLE005412.35$-$732244.3 & 0.562853 & 2450450.39412 & 19.01 & 19.49 & 19.78 \\
SMC$\_$SC6  & OGLE005216.92$-$732104.6 & 0.602526 & 2450450.04994 & 19.16 & 19.77 & 20.16 \\
SMC$\_$SC6  & OGLE005253.73$-$731858.8 & 0.635921 & 2450450.34560 & 18.92 & 19.46 & 19.80 \\
SMC$\_$SC6  & OGLE005142.15$-$731719.9 & 0.612405 & 2450450.50647 & 18.89 & 19.33 & 19.64 \\
SMC$\_$SC6  & OGLE005342.12$-$731642.7 & 0.575976 & 2450450.21173 & 19.09 & 19.64 & 20.00 \\
SMC$\_$SC6  & OGLE005308.68$-$731542.2 & 0.685501 & 2450450.45818 & 18.90 & 19.53 & 19.92 \\
SMC$\_$SC6  & OGLE005244.35$-$731258.6 & 0.570412 & 2450450.52856 & 19.48 & 20.11 & 20.31 \\
SMC$\_$SC6  & OGLE005415.82$-$731137.2 & 0.688832 & 2450450.57601 & 18.99 & 19.68 & 19.98 \\
SMC$\_$SC6  & OGLE005408.31$-$730948.2 & 0.543769 & 2450450.03629 & 19.29 & 19.61 & 20.39 \\
SMC$\_$SC6  & OGLE005312.51$-$730832.5 & 0.532397 & 2450450.12466 & 19.25 & 19.84 & 19.97 \\
SMC$\_$SC6  & OGLE005311.88$-$730832.4 & 0.600040 & 2450450.12622 & 19.06 & 19.42 & 19.97 \\
SMC$\_$SC6  & OGLE005253.76$-$730523.5 & 0.501068 & 2450450.46047 & 19.50 & 20.11 & 20.29 \\
SMC$\_$SC6  & OGLE005251.40$-$730430.7 & 0.511478 & 2450450.17757 & 19.50 & 20.04 & 20.36 \\
SMC$\_$SC6  & OGLE005326.07$-$730414.0 & 0.683569 & 2450450.26657 & 19.01 & 19.64 & 20.09 \\
SMC$\_$SC6  & OGLE005144.52$-$730150.8 & 0.524808 & 2450450.12880 & 19.36 & 19.89 & 20.25 \\
SMC$\_$SC6  & OGLE005302.56$-$730133.3 & 0.598191 & 2450450.42114 & 19.25 & 19.75 & 20.24 \\
SMC$\_$SC6  & OGLE005249.73$-$730053.2 & 0.513645 & 2450450.42402 & 19.36 & 19.82 & 20.10 \\
SMC$\_$SC6  & OGLE005318.13$-$725759.4 & 0.616994 & 2450450.34480 & 19.02 & 19.60 & 19.85 \\
SMC$\_$SC6  & OGLE005202.85$-$725707.3 & 0.798955 & 2450450.50349 & 19.13 & 20.13 & 20.52 \\
SMC$\_$SC6  & OGLE005424.14$-$725310.0 & 0.584825 & 2450450.18230 & 19.24 & 19.91 & 20.34 \\
SMC$\_$SC6  & OGLE005126.54$-$725205.3 & 0.543707 & 2450450.24597 & 19.33 & 19.69 &  --   \\
SMC$\_$SC6  & OGLE005300.26$-$725136.6 & 0.403584 & 2450450.19951 & 19.50 & 20.01 & 20.36 \\
SMC$\_$SC6  & OGLE005415.50$-$725039.2 & 0.530955 & 2450450.48345 & 19.10 & 19.65 & 20.35 \\
SMC$\_$SC6  & OGLE005428.43$-$724840.3 & 0.579410 & 2450450.24757 & 19.11 & 19.71 & 20.22 \\
SMC$\_$SC6  & OGLE005417.74$-$724755.1 & 0.611398 & 2450450.48150 & 19.16 & 20.00 & 20.50 \\
SMC$\_$SC6  & OGLE005243.99$-$724740.5 & 0.611447 & 2450450.20140 & 19.02 & 19.53 & 19.84 \\
SMC$\_$SC6  & OGLE005211.81$-$724646.2 & 0.544049 & 2450450.41321 & 19.22 & 19.71 & 20.06 \\
SMC$\_$SC6  & OGLE005202.38$-$724638.8 & 0.757914 & 2450450.45342 & 19.11 & 19.84 & 20.34 \\
SMC$\_$SC6  & OGLE005358.43$-$724538.9 & 0.553928 & 2450450.00327 & 19.30 & 19.82 &  --   \\
SMC$\_$SC6  & OGLE005139.52$-$724334.0 & 0.475422 & 2450450.04202 & 19.48 & 19.88 & 20.63 \\
SMC$\_$SC6  & OGLE005300.78$-$724245.7 & 0.567517 & 2450450.35721 & 19.33 & 20.10 & 20.67 \\
SMC$\_$SC6  & OGLE005200.56$-$724209.9 & 0.587970 & 2450450.03713 & 19.44 & 20.09 & 20.56 \\
\noalign{\vskip3pt}
\hline}

\setcounter{table}{1}
\MakeTableSepp{
l@{\hspace{4pt}}
l@{\hspace{4pt}}
c@{\hspace{4pt}}
c@{\hspace{4pt}}
c@{\hspace{4pt}}
c@{\hspace{4pt}}
c@{\hspace{4pt}}}
{12.5cm}{Continued}
{\hline
\noalign{\vskip3pt}
\multicolumn{1}{c}{Field} & \multicolumn{1}{c}{Star ID} & 
$P$ & $T_0$ & $I$ & $V$ & $B$ \\
& & [days] & [HJD] & [mag] & [mag] & [mag] \\
\noalign{\vskip3pt}
\hline
\noalign{\vskip3pt}
SMC$\_$SC6  & OGLE005142.52$-$724101.0 & 0.645529 & 2450450.10923 & 18.98 & 20.02 & 20.13 \\
SMC$\_$SC6  & OGLE005229.18$-$723902.5 & 0.612973 & 2450450.46846 & 19.50 & 20.15 & 20.68 \\
SMC$\_$SC6  & OGLE005410.32$-$723725.7 & 0.628749 & 2450450.56958 & 18.48 & 19.02 & 19.10 \\
SMC$\_$SC6  & OGLE005329.90$-$723600.3 & 0.612264 & 2450450.33708 & 19.07 &  --   &  --   \\
SMC$\_$SC6  & OGLE005236.85$-$723519.5 & 0.510831 & 2450450.13267 & 18.73 &  --   &  --   \\
SMC$\_$SC6  & OGLE005334.07$-$723508.1 & 0.637590 & 2450450.49360 & 18.81 & 19.36 & 19.78 \\
SMC$\_$SC6  & OGLE005324.12$-$723308.1 & 0.627479 & 2450450.52660 & 19.02 &  --   &  --   \\
SMC$\_$SC6  & OGLE005348.57$-$723239.5 & 0.683095 & 2450450.38499 & 19.04 & 19.68 & 20.08 \\
SMC$\_$SC7  & OGLE005450.07$-$732054.3 & 0.525835 & 2450450.23858 & 19.37 & 19.93 & 20.31 \\
SMC$\_$SC7  & OGLE005515.21$-$731918.2 & 0.589480 & 2450450.15940 & 18.74 & 19.25 & 19.61 \\
SMC$\_$SC7  & OGLE005721.02$-$731838.7 & 0.577696 & 2450450.45161 & 19.16 & 19.71 & 20.06 \\
SMC$\_$SC7  & OGLE005632.59$-$731833.6 & 0.554398 & 2450450.27031 & 19.39 & 19.96 & 20.38 \\
SMC$\_$SC7  & OGLE005613.46$-$731820.3 & 0.497593 & 2450450.19533 & 19.27 & 19.80 & 20.21 \\
SMC$\_$SC7  & OGLE005640.54$-$731620.5 & 0.629896 & 2450450.45481 & 19.09 & 19.64 & 20.04 \\
SMC$\_$SC7  & OGLE005441.17$-$731436.4 & 0.573377 & 2450450.27215 & 19.18 &  --   &  --   \\
SMC$\_$SC7  & OGLE005505.47$-$731251.9 & 0.615814 & 2450450.44364 & 19.02 & 19.60 & 19.98 \\
SMC$\_$SC7  & OGLE005612.31$-$731222.5 & 0.619898 & 2450450.14896 & 19.16 & 19.71 & 20.09 \\
SMC$\_$SC7  & OGLE005723.57$-$731136.1 & 0.562479 & 2450450.28445 & 19.03 & 19.83 & 20.11 \\
SMC$\_$SC7  & OGLE005504.67$-$731106.4 & 0.463068 & 2450450.00221 & 19.35 & 19.82 & 20.09 \\
SMC$\_$SC7  & OGLE005614.52$-$731023.8 & 0.701793 & 2450450.42131 & 19.15 & 19.91 & 20.10 \\
SMC$\_$SC7  & OGLE005616.96$-$730901.6 & 0.615658 & 2450450.13501 & 18.94 & 19.35 & 19.45 \\
SMC$\_$SC7  & OGLE005658.96$-$730850.6 & 0.641613 & 2450450.52869 & 18.68 & 19.21 & 19.73 \\
SMC$\_$SC7  & OGLE005656.87$-$730850.2 & 0.545403 & 2450450.00371 & 19.17 & 19.73 & 20.00 \\
SMC$\_$SC7  & OGLE005711.65$-$730825.3 & 0.503516 & 2450450.43009 & 19.20 & 20.05 & 19.95 \\
SMC$\_$SC7  & OGLE005451.81$-$730710.6 & 0.568647 & 2450450.19817 & 19.12 & 19.67 & 19.96 \\
SMC$\_$SC7  & OGLE005503.04$-$730548.6 & 0.483923 & 2450450.16703 & 19.51 & 20.04 & 20.46 \\
SMC$\_$SC7  & OGLE005538.66$-$730529.9 & 0.549977 & 2450450.28564 & 19.31 & 19.82 & 20.24 \\
SMC$\_$SC7  & OGLE005720.63$-$730416.8 & 0.535267 & 2450450.18999 & 19.08 & 19.63 & 19.85 \\
SMC$\_$SC7  & OGLE005451.94$-$730359.6 & 0.502497 & 2450450.41120 & 19.45 & 19.87 & 20.34 \\
SMC$\_$SC7  & OGLE005519.19$-$730344.7 & 0.532526 & 2450450.42787 & 18.94 & 19.42 & 19.66 \\
SMC$\_$SC7  & OGLE005609.42$-$730323.2 & 0.521781 & 2450450.35807 & 19.29 & 19.82 &  --   \\
SMC$\_$SC7  & OGLE005701.66$-$730321.8 & 0.562350 & 2450450.50541 & 19.39 & 19.98 & 20.51 \\
SMC$\_$SC7  & OGLE005606.56$-$730242.1 & 0.617223 & 2450450.54979 & 18.81 & 19.37 & 19.81 \\
SMC$\_$SC7  & OGLE005439.11$-$730215.5 & 0.467226 & 2450450.28134 & 19.26 & 19.70 & 19.95 \\
SMC$\_$SC7  & OGLE005723.39$-$730159.6 & 0.631765 & 2450450.41415 & 19.19 & 19.79 & 20.26 \\
SMC$\_$SC7  & OGLE005709.44$-$725959.9 & 0.651567 & 2450450.58377 & 19.29 & 19.88 & 20.31 \\
SMC$\_$SC7  & OGLE005656.16$-$725859.3 & 0.560855 & 2450450.52527 & 19.11 & 19.72 & 19.99 \\
SMC$\_$SC7  & OGLE005701.41$-$725857.4 & 0.597714 & 2450450.50969 & 19.19 & 19.97 & 20.60 \\
SMC$\_$SC7  & OGLE005518.27$-$725722.5 & 0.601763 & 2450450.20072 & 19.08 & 19.66 & 20.00 \\
SMC$\_$SC7  & OGLE005543.38$-$725605.7 & 0.570396 & 2450450.23060 & 19.17 & 19.79 & 20.09 \\
SMC$\_$SC7  & OGLE005719.62$-$725540.7 & 0.433862 & 2450450.00829 & 19.51 & 20.03 & 20.24 \\
SMC$\_$SC7  & OGLE005609.38$-$725516.2 & 0.550922 & 2450450.49719 & 19.22 & 19.77 & 20.16 \\
SMC$\_$SC7  & OGLE005534.90$-$725455.5 & 0.525181 & 2450450.00933 & 19.43 & 19.96 & 20.32 \\
SMC$\_$SC7  & OGLE005725.77$-$725438.5 & 0.605840 & 2450450.46988 & 19.23 & 19.84 & 20.23 \\
SMC$\_$SC7  & OGLE005549.04$-$725335.1 & 0.629768 & 2450450.43985 & 18.97 & 19.54 & 20.49 \\
SMC$\_$SC7  & OGLE005629.88$-$725213.0 & 0.422334 & 2450450.34766 & 18.25 & 19.07 &  --   \\
SMC$\_$SC7  & OGLE005642.92$-$725121.3 & 0.596850 & 2450450.58695 & 19.32 & 19.91 & 20.18 \\
SMC$\_$SC7  & OGLE005610.54$-$725044.6 & 0.619313 & 2450450.03600 & 18.97 & 19.62 &  --   \\
SMC$\_$SC7  & OGLE005519.94$-$725036.5 & 0.584305 & 2450450.36965 & 19.08 & 19.68 & 20.07 \\
SMC$\_$SC7  & OGLE005458.09$-$724948.9 & 0.447471 & 2450450.40267 & 19.35 & 19.74 & 20.14 \\
SMC$\_$SC7  & OGLE005530.88$-$724846.8 & 0.719676 & 2450450.34261 & 19.04 & 19.59 & 19.99 \\
SMC$\_$SC7  & OGLE005428.43$-$724840.3 & 0.579403 & 2450450.28475 & 19.16 & 19.70 &  --   \\
SMC$\_$SC7  & OGLE005701.81$-$724743.5 & 0.584233 & 2450450.57791 & 19.16 & 19.82 & 20.22 \\
SMC$\_$SC7  & OGLE005641.27$-$724723.7 & 0.492409 & 2450450.19256 & 19.69 & 20.24 & 20.82 \\
SMC$\_$SC7  & OGLE005705.34$-$724623.5 & 0.622923 & 2450450.54472 & 19.16 & 19.89 & 20.40 \\
SMC$\_$SC7  & OGLE005509.14$-$724607.5 & 0.616585 & 2450450.25617 & 19.10 & 19.68 & 20.12 \\
SMC$\_$SC7  & OGLE005559.00$-$724508.1 & 0.649027 & 2450450.19604 & 18.92 & 19.45 &  --   \\
SMC$\_$SC7  & OGLE005522.80$-$724505.4 & 0.606779 & 2450450.25089 & 19.01 & 19.63 & 20.05 \\
SMC$\_$SC7  & OGLE005714.69$-$724355.5 & 0.659379 & 2450450.50109 & 19.02 & 19.69 & 20.17 \\
SMC$\_$SC7  & OGLE005728.64$-$724135.0 & 0.602738 & 2450450.28085 & 19.03 & 19.67 & 20.87 \\
SMC$\_$SC7  & OGLE005506.01$-$723610.0 & 0.605221 & 2450450.44714 & 19.20 & 19.85 & 20.16 \\
SMC$\_$SC7  & OGLE005719.60$-$723529.9 & 0.642422 & 2450450.12457 & 19.54 & 20.31 & 20.77 \\
SMC$\_$SC7  & OGLE005649.36$-$723506.8 & 0.560811 & 2450450.27144 & 19.28 & 20.16 & 19.68 \\
SMC$\_$SC7  & OGLE005728.85$-$723454.6 & 0.416260 & 2450450.25103 & 19.72 & 20.35 & 20.62 \\
SMC$\_$SC7  & OGLE005646.16$-$723452.2 & 0.445986 & 2450450.24824 & 19.51 & 20.05 & 20.44 \\
SMC$\_$SC7  & OGLE005536.64$-$723410.8 & 0.519047 & 2450450.34160 & 19.43 & 20.05 & 20.53 \\
SMC$\_$SC7  & OGLE005440.61$-$723349.7 & 0.623934 & 2450450.33909 & 19.34 & 20.12 & 20.48 \\
SMC$\_$SC7  & OGLE005643.06$-$722650.5 & 0.631589 & 2450450.35773 & 19.27 & 20.01 & 20.40 \\
\noalign{\vskip3pt}
\hline}

\setcounter{table}{1}
\MakeTableSepp{
l@{\hspace{4pt}}
l@{\hspace{4pt}}
c@{\hspace{4pt}}
c@{\hspace{4pt}}
c@{\hspace{4pt}}
c@{\hspace{4pt}}
c@{\hspace{4pt}}}
{12.5cm}{Continued}
{\hline
\noalign{\vskip3pt}
\multicolumn{1}{c}{Field} & \multicolumn{1}{c}{Star ID} & 
$P$ & $T_0$ & $I$ & $V$ & $B$ \\
& & [days] & [HJD] & [mag] & [mag] & [mag] \\
\noalign{\vskip3pt}
\hline
\noalign{\vskip3pt}
SMC$\_$SC7  & OGLE005450.02$-$722645.0 & 0.607395 & 2450450.49414 & 19.35 & 19.96 & 20.29 \\
SMC$\_$SC7  & OGLE005714.48$-$722620.0 & 0.672639 & 2450450.15216 & 18.99 & 19.74 & 20.15 \\
SMC$\_$SC8  & OGLE005957.83$-$730647.6 & 0.447303 & 2450450.22832 & 19.44 & 19.92 & 20.26 \\
SMC$\_$SC8  & OGLE005838.13$-$730642.1 & 0.627792 & 2450450.30504 & 19.17 & 19.80 &  --   \\
SMC$\_$SC8  & OGLE005741.65$-$730440.5 & 0.524742 & 2450450.27770 & 19.14 & 19.64 &  --   \\
SMC$\_$SC8  & OGLE005723.39$-$730159.6 & 0.631767 & 2450450.35540 & 19.22 & 19.79 &  --   \\
SMC$\_$SC8  & OGLE005836.07$-$730053.0 & 0.630980 & 2450450.26619 & 19.10 & 19.79 & 20.04 \\
SMC$\_$SC8  & OGLE005835.93$-$730031.2 & 0.574874 & 2450450.48963 & 19.14 & 19.59 & 19.86 \\
SMC$\_$SC8  & OGLE005857.37$-$725948.2 & 0.620118 & 2450450.60867 & 19.04 & 19.55 & 20.51 \\
SMC$\_$SC8  & OGLE005858.64$-$725935.2 & 0.602382 & 2450450.15569 & 19.17 & 19.70 &  --   \\
SMC$\_$SC8  & OGLE005740.55$-$725726.1 & 0.624170 & 2450450.11934 & 19.23 & 19.72 & 20.18 \\
SMC$\_$SC8  & OGLE005848.70$-$725713.4 & 0.595096 & 2450450.36930 & 18.97 & 19.50 & 19.91 \\
SMC$\_$SC8  & OGLE005813.16$-$725530.0 & 0.587750 & 2450450.23035 & 18.87 & 19.34 & 19.68 \\
SMC$\_$SC8  & OGLE005725.77$-$725438.5 & 0.605842 & 2450450.47481 & 19.26 & 19.89 &  --   \\
SMC$\_$SC8  & OGLE010010.82$-$725400.0 & 0.619525 & 2450450.60344 & 18.54 & 19.12 & 19.40 \\
SMC$\_$SC8  & OGLE005832.99$-$725355.4 & 0.618296 & 2450450.54077 & 19.30 & 19.95 & 20.66 \\
SMC$\_$SC8  & OGLE005849.63$-$724943.2 & 0.611491 & 2450450.10523 & 18.96 & 19.50 & 19.95 \\
SMC$\_$SC8  & OGLE005928.02$-$724852.2 & 0.649107 & 2450450.16574 & 19.36 & 20.05 &  --   \\
SMC$\_$SC8  & OGLE005915.72$-$724831.7 & 0.552687 & 2450450.32443 & 19.23 & 19.85 & 20.12 \\
SMC$\_$SC8  & OGLE005849.95$-$724547.3 & 0.621396 & 2450450.60899 & 18.77 & 19.32 & 19.65 \\
SMC$\_$SC8  & OGLE005744.12$-$724324.8 & 0.609748 & 2450450.32304 & 19.18 & 19.80 & 20.17 \\
SMC$\_$SC8  & OGLE005752.70$-$724306.1 & 0.594998 & 2450450.33809 & 19.07 & 19.67 & 20.17 \\
SMC$\_$SC8  & OGLE005849.59$-$724211.6 & 0.654940 & 2450450.07426 & 19.06 & 19.72 & 20.14 \\
SMC$\_$SC8  & OGLE005728.64$-$724135.0 & 0.602733 & 2450450.28289 & 19.08 & 19.71 &  --   \\
SMC$\_$SC8  & OGLE005943.17$-$724047.4 & 0.466507 & 2450450.10486 & 19.46 & 19.97 & 20.61 \\
SMC$\_$SC8  & OGLE005752.74$-$723901.8 & 0.557047 & 2450450.07982 & 19.27 & 19.87 & 20.34 \\
SMC$\_$SC8  & OGLE005820.37$-$723841.4 & 0.640661 & 2450450.63266 & 18.76 & 19.36 & 19.87 \\
SMC$\_$SC8  & OGLE005737.26$-$723819.1 & 0.687028 & 2450450.41877 & 19.20 & 19.84 & 20.31 \\
SMC$\_$SC8  & OGLE005806.86$-$723811.7 & 0.614224 & 2450450.35984 & 18.89 & 19.48 & 19.89 \\
SMC$\_$SC8  & OGLE005933.82$-$723607.1 & 0.592689 & 2450450.27611 & 19.13 & 19.76 & 20.08 \\
SMC$\_$SC8  & OGLE005803.73$-$723602.0 & 0.625510 & 2450450.27711 & 19.21 & 19.89 & 20.44 \\
SMC$\_$SC8  & OGLE005728.85$-$723454.6 & 0.416258 & 2450450.25297 & 19.76 & 20.29 & 20.55 \\
SMC$\_$SC8  & OGLE005905.67$-$723050.1 & 0.564275 & 2450450.29164 & 19.21 & 19.77 & 20.17 \\
SMC$\_$SC8  & OGLE005926.80$-$722526.1 & 0.655476 & 2450450.53693 & 19.04 & 19.84 &  --   \\
SMC$\_$SC8  & OGLE005912.21$-$722210.0 & 0.618612 & 2450450.24236 & 18.61 & 19.20 & 19.53 \\
SMC$\_$SC8  & OGLE005740.36$-$721934.4 & 0.587544 & 2450450.06505 & 19.17 & 20.09 & 19.84 \\
SMC$\_$SC8  & OGLE005923.37$-$721908.0 & 0.581425 & 2450450.31135 & 19.05 &  --   & 20.19 \\
SMC$\_$SC8  & OGLE010003.88$-$721705.3 & 0.626031 & 2450450.27826 & 19.20 & 19.96 & 20.31 \\
SMC$\_$SC8  & OGLE005939.69$-$721545.8 & 0.645881 & 2450450.59242 & 19.06 & 19.69 & 20.08 \\
SMC$\_$SC8  & OGLE005942.61$-$721452.8 & 0.563973 & 2450450.49746 & 19.04 & 19.56 & 20.01 \\
SMC$\_$SC9  & OGLE010302.96$-$725901.7 & 0.603724 & 2450450.40897 & 19.12 & 19.72 & 20.02 \\
SMC$\_$SC9  & OGLE010121.94$-$725637.1 & 0.547720 & 2450450.39358 & 19.12 & 19.62 & 20.05 \\
SMC$\_$SC9  & OGLE010140.83$-$725606.1 & 0.481324 & 2450450.23152 & 19.45 & 19.98 &  --   \\
SMC$\_$SC9  & OGLE010134.33$-$725427.4 & 0.433496 & 2450450.27980 & 19.40 & 19.94 & 20.17 \\
SMC$\_$SC9  & OGLE010247.17$-$725227.1 & 0.621718 & 2450450.14189 & 19.13 & 19.68 & 20.06 \\
SMC$\_$SC9  & OGLE010301.04$-$725208.8 & 0.693484 & 2450450.48943 & 19.18 & 19.83 & 20.24 \\
SMC$\_$SC9  & OGLE010110.58$-$725042.6 & 0.646732 & 2450450.38567 & 19.09 & 19.69 & 20.06 \\
SMC$\_$SC9  & OGLE010202.86$-$725034.1 & 0.593268 & 2450450.22450 & 19.23 & 19.76 & 20.91 \\
SMC$\_$SC9  & OGLE010046.71$-$724839.8 & 0.561071 & 2450450.17836 & 18.99 & 19.54 & 20.03 \\
SMC$\_$SC9  & OGLE010135.31$-$724803.7 & 0.558498 & 2450450.45648 & 19.13 & 19.65 & 20.08 \\
SMC$\_$SC9  & OGLE010244.66$-$724613.8 & 0.542216 & 2450450.42714 & 19.10 & 19.64 & 20.16 \\
SMC$\_$SC9  & OGLE010147.55$-$724554.7 & 0.717976 & 2450450.46503 & 18.45 & 18.97 & 19.26 \\
SMC$\_$SC9  & OGLE010144.36$-$724332.6 & 0.634869 & 2450450.38200 & 19.04 & 19.60 & 19.86 \\
SMC$\_$SC9  & OGLE010218.16$-$724304.7 & 0.580739 & 2450450.39420 & 19.04 & 19.60 & 19.98 \\
SMC$\_$SC9  & OGLE010130.04$-$724210.6 & 0.590646 & 2450450.38560 & 19.40 & 19.91 & 20.20 \\
SMC$\_$SC9  & OGLE010249.81$-$724157.9 & 0.620329 & 2450450.32296 & 19.01 & 19.61 & 19.95 \\
SMC$\_$SC9  & OGLE010124.81$-$724157.0 & 0.533424 & 2450450.43280 & 19.36 & 19.90 & 20.24 \\
SMC$\_$SC9  & OGLE010219.99$-$724143.4 & 0.512806 & 2450450.30438 & 19.04 & 19.53 & 19.64 \\
SMC$\_$SC9  & OGLE010245.87$-$724107.0 & 0.567842 & 2450450.00285 & 19.17 & 19.73 & 20.17 \\
SMC$\_$SC9  & OGLE010249.16$-$723736.4 & 0.708618 & 2450450.43466 & 19.14 & 19.79 & 20.29 \\
SMC$\_$SC9  & OGLE010232.98$-$723534.4 & 0.557708 & 2450450.44018 & 19.17 & 19.67 & 20.06 \\
SMC$\_$SC9  & OGLE010042.76$-$723415.3 & 0.535404 & 2450450.12583 & 19.48 & 20.04 &  --   \\
SMC$\_$SC9  & OGLE010148.87$-$723212.1 & 0.754425 & 2450450.60292 & 18.88 & 19.25 & 19.52 \\
SMC$\_$SC9  & OGLE010102.42$-$722844.1 & 0.657289 & 2450450.38268 & 19.16 & 19.80 & 20.37 \\
SMC$\_$SC9  & OGLE010127.68$-$722700.0 & 0.630669 & 2450450.50706 & 19.33 & 20.06 & 20.38 \\
SMC$\_$SC9  & OGLE010235.05$-$722255.4 & 0.603652 & 2450450.32844 & 19.37 & 19.96 & 20.32 \\
SMC$\_$SC9  & OGLE010302.01$-$722213.0 & 0.514681 & 2450450.01568 & 19.27 & 19.91 & 20.09 \\
SMC$\_$SC9  & OGLE010125.15$-$721904.9 & 0.654753 & 2450450.25568 & 18.85 & 19.47 & 19.84 \\
SMC$\_$SC9  & OGLE010255.53$-$721643.6 & 0.593289 & 2450450.38856 & 18.82 & 19.33 & 19.74 \\
\noalign{\vskip3pt}
\hline}

\setcounter{table}{1}
\MakeTableSepp{
l@{\hspace{4pt}}
l@{\hspace{4pt}}
c@{\hspace{4pt}}
c@{\hspace{4pt}}
c@{\hspace{4pt}}
c@{\hspace{4pt}}
c@{\hspace{4pt}}}
{12.5cm}{Concluded}
{\hline
\noalign{\vskip3pt}
\multicolumn{1}{c}{Field} & \multicolumn{1}{c}{Star ID} & 
$P$ & $T_0$ & $I$ & $V$ & $B$ \\
& & [days] & [HJD] & [mag] & [mag] & [mag] \\
\noalign{\vskip3pt}
\hline
\noalign{\vskip3pt}
SMC$\_$SC9  & OGLE010326.63$-$721409.8 & 0.559307 & 2450450.32551 & 19.34 & 20.00 &  --   \\
SMC$\_$SC9  & OGLE010052.12$-$721051.7 & 0.583265 & 2450450.43190 & 18.99 & 19.52 & 19.91 \\
SMC$\_$SC9  & OGLE010045.01$-$720946.8 & 0.576859 & 2450450.07281 & 18.86 & 19.38 & 19.66 \\
SMC$\_$SC9  & OGLE010324.09$-$720545.7 & 0.637207 & 2450450.04729 & 19.26 & 19.92 & 20.27 \\
SMC$\_$SC10 & OGLE010455.18$-$725156.2 & 0.541616 & 2450450.39115 & 19.30 & 20.05 & 20.25 \\
SMC$\_$SC10 & OGLE010503.64$-$725130.2 & 0.636944 & 2450450.39096 & 18.96 & 19.60 & 20.02 \\
SMC$\_$SC10 & OGLE010343.94$-$725126.3 & 0.633037 & 2450450.10653 & 19.05 & 19.57 & 20.04 \\
SMC$\_$SC10 & OGLE010432.13$-$724958.4 & 0.629954 & 2450450.47432 & 18.74 & 19.14 &  --   \\
SMC$\_$SC10 & OGLE010338.67$-$724949.6 & 0.556086 & 2450450.47132 & 18.90 & 19.23 & 19.61 \\
SMC$\_$SC10 & OGLE010448.82$-$724945.3 & 0.650334 & 2450450.12730 & 19.06 & 19.69 & 20.13 \\
SMC$\_$SC10 & OGLE010501.92$-$724932.8 & 0.496103 & 2450450.36959 & 19.21 & 19.78 &  --   \\
SMC$\_$SC10 & OGLE010450.99$-$724700.3 & 0.488994 & 2450450.31966 & 19.25 & 19.80 & 20.26 \\
SMC$\_$SC10 & OGLE010442.62$-$724627.5 & 0.614622 & 2450450.44923 & 19.15 & 19.76 & 20.10 \\
SMC$\_$SC10 & OGLE010409.82$-$724611.9 & 0.644261 & 2450450.55242 & 19.03 & 19.59 & 19.80 \\
SMC$\_$SC10 & OGLE010553.86$-$724436.4 & 0.598127 & 2450450.47656 & 19.04 & 19.61 & 20.05 \\
SMC$\_$SC10 & OGLE010446.17$-$724232.1 & 0.573893 & 2450450.54631 & 19.14 & 19.68 & 20.22 \\
SMC$\_$SC10 & OGLE010338.90$-$724034.3 & 0.648203 & 2450450.63146 & 19.35 & 19.87 & 20.18 \\
SMC$\_$SC10 & OGLE010452.90$-$724025.9 & 0.433623 & 2450450.21677 & 19.19 & 19.57 & 19.90 \\
SMC$\_$SC10 & OGLE010523.13$-$723829.2 & 0.535858 & 2450450.09156 & 19.57 & 20.17 &  --   \\
SMC$\_$SC10 & OGLE010403.41$-$723557.5 & 0.584941 & 2450450.47549 & 19.30 & 19.86 &  --   \\
SMC$\_$SC10 & OGLE010446.99$-$723456.8 & 0.503731 & 2450450.32942 & 19.27 & 19.86 & 20.25 \\
SMC$\_$SC10 & OGLE010448.37$-$722912.1 & 0.578879 & 2450450.03632 & 19.34 & 19.91 & 20.23 \\
SMC$\_$SC10 & OGLE010515.10$-$722528.3 & 0.551041 & 2450450.27579 & 19.09 & 19.67 & 19.90 \\
SMC$\_$SC10 & OGLE010516.55$-$722526.5 & 0.455958 & 2450450.22964 & 19.50 & 20.09 & 20.52 \\
SMC$\_$SC10 & OGLE010415.01$-$722458.5 & 0.654505 & 2450450.59721 & 18.82 & 19.38 & 19.81 \\
SMC$\_$SC10 & OGLE010427.09$-$722320.1 & 0.560479 & 2450450.13678 & 19.25 & 19.80 & 20.20 \\
SMC$\_$SC10 & OGLE010431.15$-$722125.8 & 0.533258 & 2450450.23179 & 19.22 & 19.73 & 20.11 \\
SMC$\_$SC10 & OGLE010434.89$-$721722.2 & 0.578845 & 2450450.45562 & 19.40 & 20.09 & 20.40 \\
SMC$\_$SC10 & OGLE010326.63$-$721409.8 & 0.559307 & 2450450.32346 & 19.36 & 19.94 &  --   \\
SMC$\_$SC10 & OGLE010445.62$-$721035.2 & 0.681351 & 2450450.22679 & 18.99 & 19.63 & 20.03 \\
SMC$\_$SC10 & OGLE010332.65$-$720944.8 & 0.614248 & 2450450.23793 & 18.74 & 19.14 & 19.43 \\
SMC$\_$SC10 & OGLE010535.93$-$720621.6 & 0.453841 & 2450450.09247 & 19.50 & 19.93 & 20.34 \\
SMC$\_$SC10 & OGLE010446.97$-$720620.4 & 0.749592 & 2450450.50803 & 19.05 & 19.68 & 20.12 \\
SMC$\_$SC10 & OGLE010324.09$-$720545.7 & 0.637209 & 2450450.05105 & 19.25 & 19.89 &  --   \\
SMC$\_$SC10 & OGLE010603.11$-$720500.6 & 0.568160 & 2450450.21802 & 19.41 & 20.04 & 20.27 \\
SMC$\_$SC10 & OGLE010344.20$-$720407.9 & 0.632913 & 2450450.26553 & 19.00 & 19.54 & 19.87 \\
SMC$\_$SC10 & OGLE010455.87$-$720320.2 & 0.643898 & 2450450.51066 & 18.92 & 19.51 & 19.88 \\
SMC$\_$SC10 & OGLE010509.23$-$720218.0 & 0.599654 & 2450450.26605 & 18.94 & 19.55 & 19.90 \\
SMC$\_$SC10 & OGLE010411.34$-$720152.0 & 0.573859 & 2450450.01135 & 19.56 & 20.18 & 20.65 \\
SMC$\_$SC10 & OGLE010611.31$-$715732.7 & 0.643145 & 2450450.26498 & 19.00 & 19.64 & 20.04 \\
SMC$\_$SC10 & OGLE010403.48$-$715721.6 & 0.616891 & 2450450.33076 & 19.30 & 19.91 & 20.38 \\
SMC$\_$SC10 & OGLE010359.90$-$715720.0 & 0.611870 & 2450450.30397 & 18.62 & 19.16 & 19.50 \\
SMC$\_$SC11 & OGLE010800.59$-$730739.5 & 0.623917 & 2450450.28937 & 19.06 &  --   &  --   \\
SMC$\_$SC11 & OGLE010847.54$-$730521.1 & 0.568065 & 2450450.34059 & 19.33 & 19.94 &  --   \\
SMC$\_$SC11 & OGLE010842.27$-$730411.2 & 0.755208 & 2450450.44509 & 19.00 & 19.60 & 20.06 \\
SMC$\_$SC11 & OGLE010834.94$-$730005.6 & 0.557271 & 2450450.11485 & 19.10 & 19.67 &  --   \\
SMC$\_$SC11 & OGLE010725.95$-$725941.5 & 0.580738 & 2450450.30387 & 18.93 & 19.58 & 19.76 \\
SMC$\_$SC11 & OGLE010823.35$-$725813.9 & 0.785200 & 2450450.32256 & 18.78 & 19.46 & 19.98 \\
SMC$\_$SC11 & OGLE010640.65$-$725637.3 & 0.574302 & 2450450.04248 & 19.15 & 19.69 &  --   \\
SMC$\_$SC11 & OGLE010627.06$-$725434.5 & 0.649536 & 2450450.53118 & 18.69 & 19.30 &  --   \\
SMC$\_$SC11 & OGLE010653.16$-$725406.9 & 0.562441 & 2450450.03668 & 19.14 & 19.63 &  --   \\
SMC$\_$SC11 & OGLE010836.97$-$725321.6 & 0.619422 & 2450450.22081 & 19.16 & 19.76 &  --   \\
SMC$\_$SC11 & OGLE010909.88$-$724646.4 & 0.658584 & 2450450.35160 & 18.99 & 19.60 &  --   \\
SMC$\_$SC11 & OGLE010909.66$-$724149.8 & 0.565590 & 2450450.21404 & 18.95 & 19.54 & 19.85 \\
SMC$\_$SC11 & OGLE010654.39$-$724145.5 & 0.578431 & 2450450.50201 & 19.18 & 19.71 & 20.04 \\
SMC$\_$SC11 & OGLE010648.19$-$723553.4 & 0.525160 & 2450450.22892 & 19.34 & 19.94 & 20.21 \\
SMC$\_$SC11 & OGLE010633.08$-$723544.9 & 0.627022 & 2450450.39759 & 19.32 & 19.99 & 20.45 \\
SMC$\_$SC11 & OGLE010804.44$-$723349.3 & 0.646477 & 2450450.33238 & 18.88 & 19.54 & 19.91 \\
SMC$\_$SC11 & OGLE010846.41$-$723243.2 & 0.579876 & 2450450.01153 & 19.13 & 19.73 & 20.18 \\
SMC$\_$SC11 & OGLE010728.57$-$723052.9 & 0.575224 & 2450450.54781 & 19.00 & 19.58 & 19.84 \\
SMC$\_$SC11 & OGLE010838.76$-$722758.0 & 0.622726 & 2450450.53749 & 18.87 & 19.32 & 19.58 \\
SMC$\_$SC11 & OGLE010655.60$-$722754.5 & 0.510401 & 2450450.05283 & 19.12 & 19.55 & 19.85 \\
SMC$\_$SC11 & OGLE010742.39$-$722352.8 & 0.529878 & 2450450.08817 & 18.89 & 19.43 & 19.71 \\
SMC$\_$SC11 & OGLE010848.72$-$722342.9 & 0.626908 & 2450450.36074 & 19.14 & 19.71 & 20.00 \\
SMC$\_$SC11 & OGLE010818.39$-$722315.5 & 0.561962 & 2450450.46985 & 19.23 & 19.71 &  --   \\
SMC$\_$SC11 & OGLE010721.87$-$721825.9 & 0.672973 & 2450450.37001 & 18.99 & 19.62 & 19.84 \\
SMC$\_$SC11 & OGLE010802.60$-$721600.8 & 0.629378 & 2450450.55957 & 18.84 & 19.43 & 19.82 \\
SMC$\_$SC11 & OGLE010734.49$-$721429.8 & 0.646715 & 2450450.38464 & 18.98 & 19.60 & 20.00 \\
SMC$\_$SC11 & OGLE010838.27$-$721333.3 & 0.578293 & 2450450.36868 & 19.11 & 19.68 & 20.05 \\
\noalign{\vskip3pt}
\hline}

\renewcommand{\arraystretch}{.93}
\renewcommand{\TableFont}{\scriptsize}
\setcounter{table}{2}
\MakeTableSepp{
l@{\hspace{4pt}}
l@{\hspace{4pt}}
c@{\hspace{4pt}}
c@{\hspace{4pt}}
c@{\hspace{4pt}}
c@{\hspace{4pt}}
c@{\hspace{4pt}}}
{12.5cm}{c-type RR Lyrae stars from the SMC}
{\hline
\noalign{\vskip3pt}
\multicolumn{1}{c}{Field} & \multicolumn{1}{c}{Star ID} & 
$P$ & $T_0$ & $I$ & $V$ & $B$ \\
& & [days] & [HJD] & [mag] & [mag] & [mag] \\
\noalign{\vskip3pt}
\hline
\noalign{\vskip3pt}
SMC$\_$SC1  & OGLE003908.31$-$735426.8 & 0.281006 & 2450450.20606 & 18.99 & 19.32 &  --   \\
SMC$\_$SC1  & OGLE003744.55$-$734750.5 & 0.368667 & 2450450.23920 & 19.15 & 19.61 & 19.83 \\
SMC$\_$SC1  & OGLE003737.93$-$734706.2 & 0.352683 & 2450450.33501 & 19.32 & 19.77 & 20.04 \\
SMC$\_$SC1  & OGLE003853.80$-$732328.0 & 0.345051 & 2450450.27801 & 19.38 & 19.83 & 20.15 \\
SMC$\_$SC1  & OGLE003706.77$-$730551.2 & 0.293459 & 2450450.06673 & 19.20 & 19.43 & 19.64 \\
SMC$\_$SC2  & OGLE003946.61$-$733332.6 & 0.309804 & 2450450.17994 & 19.42 & 19.85 & 20.44 \\
SMC$\_$SC2  & OGLE004007.32$-$731338.8 & 0.343159 & 2450450.22418 & 19.44 & 19.94 & 20.27 \\
SMC$\_$SC2  & OGLE003934.35$-$730433.9 & 0.351432 & 2450450.11490 & 19.10 & 19.56 &  --   \\
SMC$\_$SC3  & OGLE004330.74$-$733416.0 & 0.320723 & 2450450.20162 & 19.29 & 19.72 & 19.94 \\
SMC$\_$SC3  & OGLE004444.50$-$732833.1 & 0.370527 & 2450450.20609 & 19.06 & 19.79 & 20.63 \\
SMC$\_$SC3  & OGLE004245.08$-$730939.7 & 0.311416 & 2450450.18060 & 19.27 & 19.64 & 19.85 \\
SMC$\_$SC3  & OGLE004322.30$-$725905.5 & 0.346199 & 2450450.27457 & 19.13 & 19.56 & 19.82 \\
SMC$\_$SC3  & OGLE004327.25$-$724542.4 & 0.344124 & 2450450.23623 & 19.20 & 19.70 & 19.92 \\
SMC$\_$SC4  & OGLE004808.56$-$731003.6 & 0.326195 & 2450450.07486 & 19.28 & 19.75 & 20.04 \\
SMC$\_$SC4  & OGLE004624.67$-$730050.3 & 0.357261 & 2450450.08049 & 18.93 & 19.57 & 20.13 \\
SMC$\_$SC4  & OGLE004744.89$-$725530.1 & 0.297777 & 2450450.00931 & 19.18 & 19.51 & 19.79 \\
SMC$\_$SC4  & OGLE004633.73$-$725253.8 & 0.302182 & 2450450.19175 & 19.50 & 19.93 & 20.21 \\
SMC$\_$SC4  & OGLE004631.02$-$724658.8 & 0.308220 & 2450450.28206 & 19.24 & 19.63 & 19.86 \\
SMC$\_$SC5  & OGLE004848.01$-$732714.6 & 0.345412 & 2450450.27837 & 19.11 & 19.41 & 19.64 \\
SMC$\_$SC5  & OGLE004845.41$-$730826.4 & 0.372181 & 2450450.15355 & 19.36 & 19.87 & 20.31 \\
SMC$\_$SC5  & OGLE005115.64$-$724739.2 & 0.280936 & 2450450.18512 & 19.16 & 19.82 & 19.82 \\
SMC$\_$SC5  & OGLE004902.13$-$724513.6 & 0.310884 & 2450450.24968 & 19.18 & 19.69 & 20.17 \\
SMC$\_$SC5  & OGLE004840.40$-$724402.4 & 0.354180 & 2450450.16695 & 19.32 & 19.79 &  --   \\
SMC$\_$SC6  & OGLE005155.37$-$724909.5 & 0.363936 & 2450450.06872 & 19.39 & 19.93 & 20.15 \\
SMC$\_$SC7  & OGLE005556.74$-$732133.6 & 0.349140 & 2450450.24051 & 19.38 &  --   &  --   \\
SMC$\_$SC7  & OGLE005503.23$-$731907.8 & 0.313536 & 2450450.24977 & 19.27 & 19.62 & 19.86 \\
SMC$\_$SC7  & OGLE005609.24$-$731504.8 & 0.277551 & 2450450.17346 & 18.78 & 19.31 & 19.65 \\
SMC$\_$SC7  & OGLE005601.69$-$730338.5 & 0.427773 & 2450450.20740 & 18.79 & 19.30 & 19.70 \\
SMC$\_$SC7  & OGLE005621.27$-$730046.9 & 0.351118 & 2450450.18641 & 19.24 & 19.71 & 19.92 \\
SMC$\_$SC7  & OGLE005633.68$-$725700.0 & 0.441252 & 2450450.25788 & 19.11 & 19.65 &  --   \\
SMC$\_$SC7  & OGLE005618.62$-$725209.7 & 0.370889 & 2450450.14595 & 18.92 & 19.36 & 19.64 \\
SMC$\_$SC7  & OGLE005725.22$-$724432.7 & 0.337364 & 2450450.27667 & 18.88 & 19.53 & 19.95 \\
SMC$\_$SC7  & OGLE005527.97$-$724136.8 & 0.371404 & 2450450.00957 & 19.22 & 19.75 & 20.03 \\
SMC$\_$SC7  & OGLE005601.43$-$724026.2 & 0.283717 & 2450450.16025 & 19.58 &  --   &  --   \\
SMC$\_$SC7  & OGLE005451.72$-$723850.4 & 0.277966 & 2450450.00890 & 19.03 & 19.31 &  --   \\
SMC$\_$SC8  & OGLE010029.57$-$725454.2 & 0.343212 & 2450450.33141 & 19.41 & 19.92 & 20.21 \\
SMC$\_$SC8  & OGLE005853.75$-$725335.2 & 0.371910 & 2450450.09010 & 19.26 & 19.75 & 20.13 \\
SMC$\_$SC8  & OGLE005913.99$-$724719.4 & 0.313798 & 2450450.30113 & 19.44 & 19.89 & 20.24 \\
SMC$\_$SC8  & OGLE005754.46$-$723248.2 & 0.283746 & 2450450.13031 & 18.52 & 19.29 & 20.11 \\
SMC$\_$SC8  & OGLE005913.24$-$722418.6 & 0.378142 & 2450450.07666 & 19.33 & 19.77 & 20.37 \\
SMC$\_$SC9  & OGLE010151.77$-$730020.1 & 0.385044 & 2450450.04714 & 19.37 &  --   &  --   \\
SMC$\_$SC9  & OGLE010029.57$-$725454.2 & 0.343216 & 2450450.32098 & 19.47 & 19.91 &  --   \\
SMC$\_$SC9  & OGLE010246.69$-$725030.1 & 0.262593 & 2450450.01799 & 18.80 & 19.09 & 19.42 \\
SMC$\_$SC9  & OGLE010316.37$-$724816.2 & 0.370538 & 2450450.13873 & 19.34 & 19.83 & 19.79 \\
SMC$\_$SC9  & OGLE010220.67$-$723753.6 & 0.261244 & 2450450.16592 & 19.61 & 19.95 &  --   \\
SMC$\_$SC9  & OGLE010256.60$-$723431.0 & 0.445955 & 2450450.03397 & 19.03 & 19.56 & 19.91 \\
SMC$\_$SC9  & OGLE010232.82$-$723147.1 & 0.347329 & 2450450.25602 & 19.11 & 19.58 & 19.71 \\
SMC$\_$SC9  & OGLE010231.17$-$722134.0 & 0.346288 & 2450450.20236 & 19.25 & 19.73 & 20.04 \\
SMC$\_$SC9  & OGLE010148.62$-$721836.8 & 0.368215 & 2450450.00787 & 19.22 & 19.66 & 19.95 \\
SMC$\_$SC9  & OGLE010245.99$-$721132.7 & 0.294040 & 2450450.16386 & 19.32 & 19.73 & 19.92 \\
SMC$\_$SC9  & OGLE010249.92$-$720947.1 & 0.316409 & 2450450.14857 & 19.38 & 19.82 & 20.10 \\
SMC$\_$SC10 & OGLE010316.37$-$724816.2 & 0.370556 & 2450450.12805 & 19.42 &  --   &  --   \\
SMC$\_$SC10 & OGLE010349.04$-$724400.1 & 0.369181 & 2450450.10959 & 19.22 & 19.72 & 20.09 \\
SMC$\_$SC10 & OGLE010453.59$-$723756.0 & 0.339862 & 2450450.14489 & 19.27 & 19.75 & 20.12 \\
SMC$\_$SC11 & OGLE010647.29$-$723053.7 & 0.362757 & 2450450.22340 & 19.25 & 19.73 & 20.07 \\
SMC$\_$SC11 & OGLE010850.08$-$723030.3 & 0.355358 & 2450450.19533 & 19.20 & 19.72 & 19.89 \\
SMC$\_$SC11 & OGLE010731.44$-$721745.7 & 0.345769 & 2450450.19988 & 19.41 & 19.89 & 20.20 \\
SMC$\_$SC11 & OGLE010718.63$-$721723.0 & 0.358960 & 2450450.24852 & 19.25 & 19.75 & 20.10 \\
\noalign{\vskip3pt}
\hline}

More parameters (period errors, Fourier parameters of the light curve 
decomposition, {\it IVB}-band amplitudes) are available in the electronic form 
from the OGLE {\sc Internet} archive: 

\begin{center}
{\it http://www.astrouw.edu.pl/\~{}ogle/} \\ {\it
ftp://sirius.astrouw.edu.pl/ogle/ogle2/var\_stars/smc/rrlyr/}\\
\end{center}
or its US mirror
\begin{center}
{\it http://bulge.princeton.edu/\~{}ogle/}\\ {\it
ftp://bulge.princeton.edu/ogle/ogle2/var\_stars/smc/rrlyr/}\\
\end{center}

\noindent
Also individual {\it BVI} observations of all objects and finding charts are 
included. 

Tables~2 and 3 contain together 536 entries but only 514 objects, because 22 
stars were detected twice -- in the overlapping regions of adjacent fields. We 
decided not to remove twice-detected RR~Lyr stars from the final list, because 
their measurements are independent in both fields and can be used for testing 
quality of data and completeness of the sample. In Table~6 we provide 
cross-reference list to identify stars in the overlapping regions. 

\begin{figure}[htb]
\centerline{\includegraphics[bb=35 350 570 750, width=11cm]{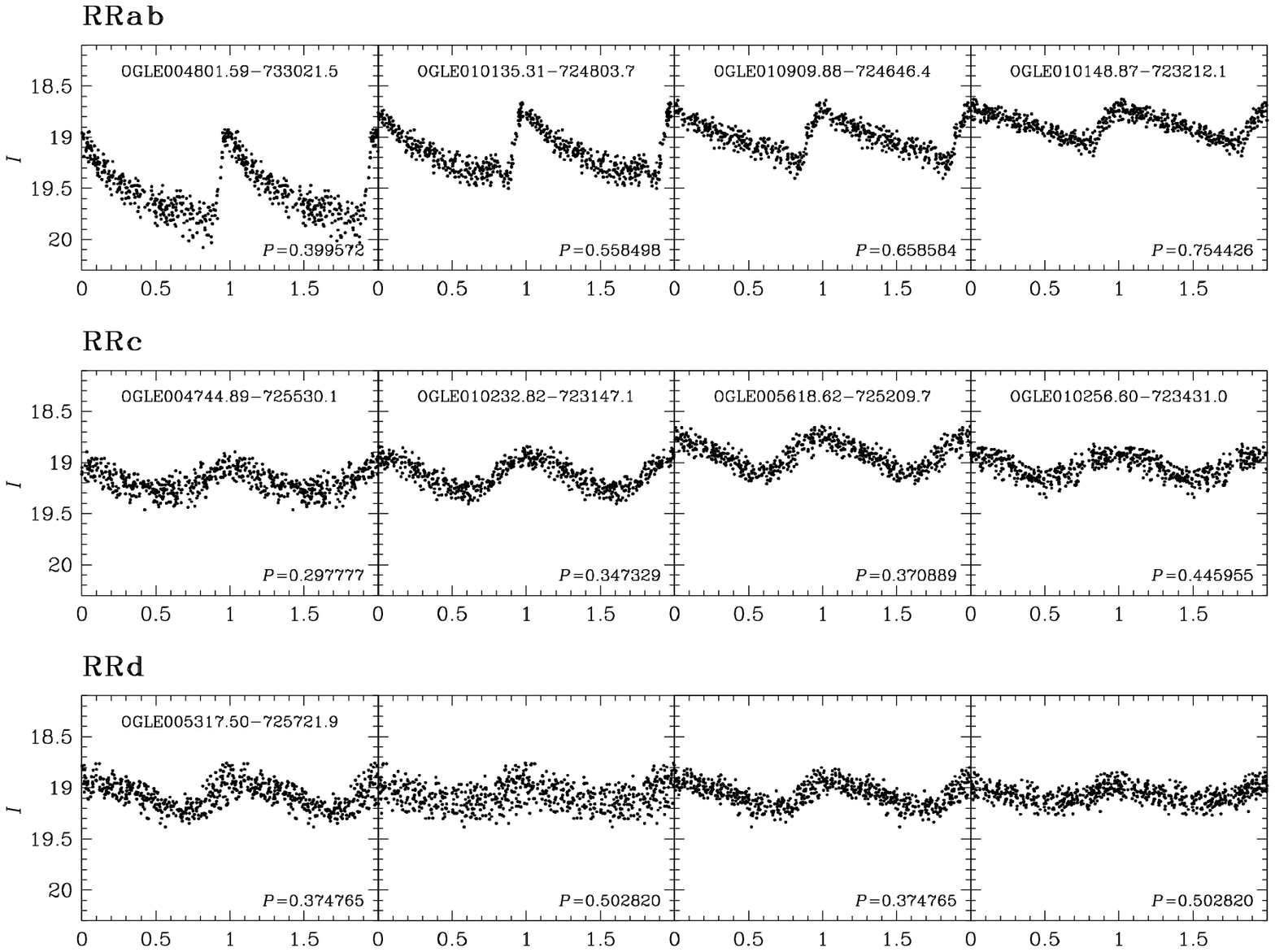}}
\FigCap{Exemplary light curves of RR Lyr stars from the SMC. In the top row 
light curves of four typical RRab stars arranged according to the periods are 
presented. In the middle row a sample of four RRc stars is presented. Bottom 
row shows the light curves of an exemplary RRd variable -- original 
photometric data folded with the shorter and longer periods, and light curves 
of each mode after subtraction of the other period variability.} 
\end{figure} 
In Fig.~2 we present {\it I}-band light curves of a~few typical RRab, RRc 
stars and one RRd star. All the diagrams have the same magnitude range to 
compare the amplitudes and brightness of the stars. The light curves are 
arranged according to the periods. 

\Subsection{Double-Mode RR Lyr Stars}
RRd stars discovered in the SMC are listed in Table~4. 59 such objects were 
detected but 57 of them are unique. Two stars are located in the overlapping 
regions between fields and they were discovered independently. Table~4 lists 
the field name, star ID, period of the first overtone, period of the 
fundamental mode, period ratio, and finally $IVB$ intensity-mean magnitudes. 
Other parameters, individual {\it BVI} measurements and finding charts are 
available from the OGLE {\sc Internet} archive. 
\renewcommand{\arraystretch}{0.88}
\renewcommand{\TableFont}{\scriptsize}
\setcounter{table}{3}
\MakeTableSep{
l@{\hspace{4pt}}
l@{\hspace{4pt}}
c@{\hspace{4pt}}
c@{\hspace{4pt}}
c@{\hspace{4pt}}
c@{\hspace{4pt}}
c@{\hspace{4pt}}
c@{\hspace{4pt}}}
{12.5cm}{d-type RR Lyrae stars from the SMC}
{\hline
\noalign{\vskip3pt}
\multicolumn{1}{c}{Field} & \multicolumn{1}{c}{Star ID} & 
$P_1$ & $P_0$ & $P_1/P_0$ & $I$ & $V$ & $B$ \\
& & [days] & [days] & & [mag] & [mag] & [mag] \\
\noalign{\vskip3pt}
\hline
\noalign{\vskip3pt}
SMC$\_$SC1  & OGLE003638.62$-$734034.2 & 0.361795 & 0.485969 & 0.744482 & 18.94 & 19.41 & 19.66 \\
SMC$\_$SC1  & OGLE003659.06$-$733452.1 & 0.372989 & 0.500389 & 0.745398 & 19.05 & 19.50 & 19.76 \\
SMC$\_$SC1  & OGLE003703.18$-$733216.6 & 0.394705 & 0.529477 & 0.745462 & 19.13 & 19.64 & 20.01 \\
SMC$\_$SC1  & OGLE003803.07$-$732829.2 & 0.380053 & 0.509537 & 0.745879 & 19.22 & 19.73 & 20.04 \\
SMC$\_$SC1  & OGLE003619.43$-$732553.0 & 0.419186 & 0.561512 & 0.746531 & 18.97 & 19.55 &   --   \\
SMC$\_$SC1  & OGLE003838.05$-$731453.7 & 0.410901 & 0.550465 & 0.746462 & 18.94 & 19.53 & 19.83 \\
SMC$\_$SC1  & OGLE003901.05$-$730751.1 & 0.397228 & 0.532444 & 0.746047 & 19.22 & 19.82 &   --   \\
SMC$\_$SC1  & OGLE003733.92$-$730607.7 & 0.364203 & 0.489339 & 0.744275 & 19.23 & 19.71 & 20.02 \\
SMC$\_$SC2  & OGLE004228.21$-$734532.0 & 0.424805 & 0.569703 & 0.745661 & 18.94 &   --   &   --   \\
SMC$\_$SC2  & OGLE004105.53$-$734149.6 & 0.387770 & 0.520756 & 0.744629 & 19.07 & 19.54 & 19.85 \\
SMC$\_$SC2  & OGLE004222.96$-$733223.8 & 0.364102 & 0.488706 & 0.745034 & 19.29 & 19.82 & 20.13 \\
SMC$\_$SC2  & OGLE004203.77$-$732544.8 & 0.378924 & 0.508841 & 0.744680 & 19.36 & 19.84 & 20.28 \\
SMC$\_$SC2  & OGLE004126.10$-$731644.9 & 0.379489 & 0.509092 & 0.745422 & 19.33 & 19.92 & 20.25 \\
SMC$\_$SC2  & OGLE004214.47$-$731644.6 & 0.379294 & 0.508928 & 0.745281 & 18.95 & 19.45 & 19.74 \\
SMC$\_$SC2  & OGLE004051.20$-$731417.8 & 0.375345 & 0.503322 & 0.745736 & 19.09 & 19.64 & 19.90 \\
SMC$\_$SC2  & OGLE004004.29$-$730044.3 & 0.367739 & 0.493854 & 0.744632 & 19.30 & 19.82 & 20.09 \\
SMC$\_$SC2  & OGLE004038.76$-$730023.5 & 0.367415 & 0.493166 & 0.745012 & 19.39 & 20.00 & 20.41 \\
SMC$\_$SC3  & OGLE004222.96$-$733223.8 & 0.364102 & 0.488706 & 0.745034 & 19.30 & 19.86 &   --   \\
SMC$\_$SC3  & OGLE004253.47$-$732750.6 & 0.354895 & 0.476748 & 0.744408 & 19.14 & 19.53 & 19.75 \\
SMC$\_$SC3  & OGLE004348.60$-$732434.9 & 0.390466 & 0.523846 & 0.745382 & 19.17 & 19.60 & 19.94 \\
SMC$\_$SC3  & OGLE004335.78$-$732058.5 & 0.423200 & 0.566785 & 0.746667 & 18.89 & 19.36 & 19.68 \\
SMC$\_$SC3  & OGLE004358.07$-$730157.7 & 0.362730 & 0.486752 & 0.745204 & 19.20 & 19.58 & 19.83 \\
SMC$\_$SC3  & OGLE004251.11$-$724804.4 & 0.379104 & 0.508353 & 0.745750 & 19.18 & 19.67 & 19.95 \\
SMC$\_$SC3  & OGLE004230.85$-$724651.0 & 0.370284 & 0.496839 & 0.745281 & 19.05 & 19.45 &   --   \\
SMC$\_$SC4  & OGLE004615.67$-$731940.4 & 0.360343 & 0.483994 & 0.744520 & 19.34 & 19.90 & 20.36 \\
SMC$\_$SC4  & OGLE004745.68$-$731516.2 & 0.402812 & 0.540698 & 0.744985 & 19.20 & 19.75 & 20.11 \\
SMC$\_$SC4  & OGLE004817.90$-$723921.2 & 0.395636 & 0.530727 & 0.745461 & 19.12 & 19.65 &   --   \\
SMC$\_$SC5  & OGLE004949.44$-$731703.1 & 0.426080 & 0.571101 & 0.746068 & 19.17 & 19.64 & 19.85 \\
SMC$\_$SC5  & OGLE005017.68$-$730910.6 & 0.382452 & 0.513099 & 0.745378 & 19.14 & 19.58 & 19.93 \\
SMC$\_$SC5  & OGLE005118.41$-$730502.2 & 0.435088 & 0.582519 & 0.746907 & 18.83 & 19.32 & 19.62 \\
SMC$\_$SC5  & OGLE005049.46$-$730030.1 & 0.373855 & 0.501734 & 0.745127 & 19.13 & 19.72 & 19.96 \\
SMC$\_$SC5  & OGLE005023.79$-$724047.7 & 0.364916 & 0.490275 & 0.744309 & 19.33 & 19.79 &   --   \\
SMC$\_$SC6  & OGLE005342.24$-$732514.1 & 0.363768 & 0.488306 & 0.744959 & 19.34 & 19.84 & 20.21 \\
SMC$\_$SC6  & OGLE005155.91$-$732428.5 & 0.378633 & 0.508443 & 0.744692 & 19.16 & 19.81 & 19.94 \\
SMC$\_$SC6  & OGLE005152.03$-$730348.6 & 0.407388 & 0.545692 & 0.746553 & 19.22 & 19.72 & 19.97 \\
SMC$\_$SC6  & OGLE005317.50$-$725721.9 & 0.374765 & 0.502820 & 0.745327 & 19.07 & 19.63 & 19.92 \\
SMC$\_$SC6  & OGLE005401.42$-$724728.7 & 0.378620 & 0.507628 & 0.745862 & 19.34 & 19.99 & 20.36 \\
SMC$\_$SC6  & OGLE005237.14$-$724651.2 & 0.372969 & 0.500557 & 0.745107 & 19.61 & 20.01 & 19.69 \\
SMC$\_$SC6  & OGLE005430.51$-$724240.9 & 0.366688 & 0.492198 & 0.745002 & 18.99 & 19.48 & 19.75 \\
SMC$\_$SC6  & OGLE005346.56$-$723637.1 & 0.388001 & 0.519872 & 0.746339 & 19.14 & 19.61 & 19.89 \\
SMC$\_$SC7  & OGLE005449.66$-$724259.5 & 0.427425 & 0.573399 & 0.745423 & 18.94 & 19.45 & 19.84 \\
SMC$\_$SC7  & OGLE005628.74$-$724247.9 & 0.392275 & 0.525936 & 0.745862 & 19.05 &   --   &   --   \\
SMC$\_$SC7  & OGLE005430.51$-$724240.9 & 0.366691 & 0.492209 & 0.744990 & 19.00 & 19.46 &   --   \\
SMC$\_$SC7  & OGLE005453.68$-$724150.7 & 0.392306 & 0.526165 & 0.745595 & 19.12 & 19.64 & 20.01 \\
SMC$\_$SC7  & OGLE005724.34$-$724031.3 & 0.396128 & 0.531318 & 0.745558 & 19.16 & 19.81 & 20.27 \\
SMC$\_$SC8  & OGLE005948.83$-$730107.6 & 0.382381 & 0.512899 & 0.745529 & 19.34 & 19.88 & 20.18 \\
SMC$\_$SC8  & OGLE010016.23$-$724726.9 & 0.369094 & 0.495560 & 0.744802 & 19.25 & 19.75 & 19.99 \\
SMC$\_$SC8  & OGLE005906.43$-$722145.3 & 0.367714 & 0.493503 & 0.745109 & 19.32 & 19.87 & 20.27 \\
SMC$\_$SC9  & OGLE010233.77$-$724728.8 & 0.366232 & 0.491380 & 0.745314 & 19.18 & 19.66 &   --   \\
SMC$\_$SC9  & OGLE010038.72$-$724455.6 & 0.370112 & 0.496828 & 0.744950 & 19.20 & 19.63 & 19.89 \\
SMC$\_$SC9  & OGLE010212.03$-$724109.6 & 0.374545 & 0.502773 & 0.744958 & 19.16 & 19.66 &   --   \\
SMC$\_$SC9  & OGLE010301.09$-$722910.9 & 0.364565 & 0.489423 & 0.744888 & 19.41 & 19.86 & 19.98 \\
SMC$\_$SC9  & OGLE010102.15$-$721554.4 & 0.365668 & 0.491466 & 0.744035 & 19.14 & 19.61 & 19.83 \\
SMC$\_$SC10 & OGLE010558.62$-$723119.0 & 0.364133 & 0.488757 & 0.745018 & 18.86 & 19.24 & 19.55 \\
SMC$\_$SC10 & OGLE010334.17$-$722141.0 & 0.376754 & 0.505382 & 0.745484 & 19.16 & 19.70 & 19.93 \\
SMC$\_$SC11 & OGLE010902.50$-$725018.9 & 0.371349 & 0.498079 & 0.745562 & 19.42 & 19.93 &   --   \\
SMC$\_$SC11 & OGLE010728.50$-$724527.6 & 0.366752 & 0.492110 & 0.745264 & 19.53 & 20.05 & 20.42 \\
SMC$\_$SC11 & OGLE010632.81$-$722035.0 & 0.359977 & 0.483701 & 0.744214 & 19.29 & 19.79 & 20.11 \\
SMC$\_$SC11 & OGLE010721.27$-$721939.9 & 0.370842 & 0.498288 & 0.744232 & 19.38 & 19.90 & 20.28 \\
\noalign{\vskip3pt}
\hline}

\Subsection{Other Stars}
Table~5 presents a list of 19 detected short-period variables and a few other 
variable stars. The table contains 20 entries because one object was 
identified twice in the overlapping region of fields SMC\_SC5 and SMC\_SC6. 
Consecutive columns of Table~5 represent the same data as in Tables~2 and~3. 
Most stars from the list are probably $\delta$~Sct stars. Two stars: 
OGLE005542.36--732105.3 and OGLE010118.99--723323.8, are double-mode 
pulsators, with the period ratios 0.76259 and 0.77098, respectively. In 
Table~5 we present longer periods of both stars. 
\renewcommand{\arraystretch}{1}
\renewcommand{\TableFont}{\footnotesize}
\setcounter{table}{4}
\MakeTable{
l@{\hspace{4pt}}
l@{\hspace{4pt}}
c@{\hspace{4pt}}
c@{\hspace{4pt}}
c@{\hspace{4pt}}
c@{\hspace{4pt}}
c@{\hspace{4pt}}}
{12.5cm}{Other variable stars from the SMC}
{\hline
\noalign{\vskip3pt}
\multicolumn{1}{c}{Field} & \multicolumn{1}{c}{Star ID} & 
$P$ & $T_0$ & $I$ & $V$ & $B$ \\
& & [days] & [HJD] & [mag] & [mag] & [mag] \\
\noalign{\vskip3pt}
\hline
\noalign{\vskip3pt}
SMC$\_$SC2  & OGLE004157.65$-$730119.9 & 0.203325 & 2450450.06205 & 19.26 & 19.82 & 20.26 \\
SMC$\_$SC2  & OGLE004115.08$-$725407.8 & 0.251083 & 2450450.23158 & 19.44 & 19.78 &  --   \\
SMC$\_$SC4  & OGLE004616.17$-$731416.1 & 0.503192 & 2450450.06849 & 18.26 & 19.02 & 19.45 \\
SMC$\_$SC5  & OGLE005008.48$-$725916.5 & 0.252511 & 2450450.12714 & 18.86 & 19.49 & 19.86 \\
SMC$\_$SC5  & OGLE005135.18$-$724438.1 & 0.238996 & 2450450.05630 & 19.07 & 19.69 &  --   \\
SMC$\_$SC6  & OGLE005424.16$-$730529.4 & 0.211744 & 2450450.09308 & 18.94 & 19.44 & 19.79 \\
SMC$\_$SC6  & OGLE005259.56$-$725605.3 & 0.356721 & 2450450.18734 & 18.27 & 19.28 & 19.69 \\
SMC$\_$SC6  & OGLE005353.94$-$725207.7 & 0.193607 & 2450450.08955 & 19.23 & 19.71 & 21.18 \\
SMC$\_$SC6  & OGLE005145.04$-$724446.5 & 0.232651 & 2450450.16994 & 19.12 & 19.67 & 20.15 \\
SMC$\_$SC6  & OGLE005135.18$-$724438.1 & 0.238996 & 2450450.08302 & 19.11 & 19.66 & 20.22 \\
SMC$\_$SC7  & OGLE005542.36$-$732105.3 & 0.341130 & 2450450.04449 & 18.88 & 19.37 & 19.75 \\
SMC$\_$SC7  & OGLE005456.26$-$730814.3 & 0.237242 & 2450450.06554 & 18.95 & 19.51 & 19.74 \\
SMC$\_$SC7  & OGLE005507.46$-$724434.0 & 0.570419 & 2450450.40706 & 18.38 & 18.75 &  --   \\
SMC$\_$SC7  & OGLE005653.43$-$724340.6 & 0.227619 & 2450450.00403 & 18.49 &  --   &  --   \\
SMC$\_$SC7  & OGLE005521.33$-$723751.8 & 0.357436 & 2450450.12745 & 18.50 & 19.01 & 19.47 \\
SMC$\_$SC7  & OGLE005558.04$-$722633.2 & 0.252600 & 2450450.17328 & 19.18 &  --   & 20.28 \\
SMC$\_$SC8  & OGLE005848.58$-$724941.6 & 0.234350 & 2450450.21723 & 18.99 & 19.44 & 19.77 \\
SMC$\_$SC8  & OGLE005739.23$-$724449.5 & 0.161751 & 2450450.01635 & 19.33 & 19.83 & 20.30 \\
SMC$\_$SC8  & OGLE005740.47$-$721413.1 & 0.226250 & 2450450.16643 & 18.93 & 19.39 & 19.77 \\
SMC$\_$SC9  & OGLE010118.99$-$723323.8 & 0.279555 & 2450450.02007 & 19.09 & 19.64 & 19.98 \\
\noalign{\vskip3pt}
\hline}

\Section{Basic Parameters of RR Lyr Stars}
\Subsection{Period Distribution}
Distribution of periods of RR~Lyr stars is an indicator of the morphology of 
the horizontal branch of the population of old, metal-weak stars. The final 
period search was carried out with program {\sc Tatry} by Schwarzenberg-Czerny 
(2002, private communication). The program works in two steps. First, it 
calculates a~periodogram and identifies its peaks. Then the corresponding peak 
frequencies are examined in detail to find the best one and its corresponding 
ephemeris. {\sc Tatry} provides also reliable estimation of period errors, 
what we could check comparing periods of stars located in the overlapping 
regions of the adjacent fields. Accuracy of periods for the vast majority of 
the RR~Lyr stars is better than $10^{-5}$ days. The period errors for each 
star are included in the OGLE {\sc Internet} archive data. 

\begin{figure}[htb]
\centerline{\includegraphics[bb=30 40 570 430, width=11cm]{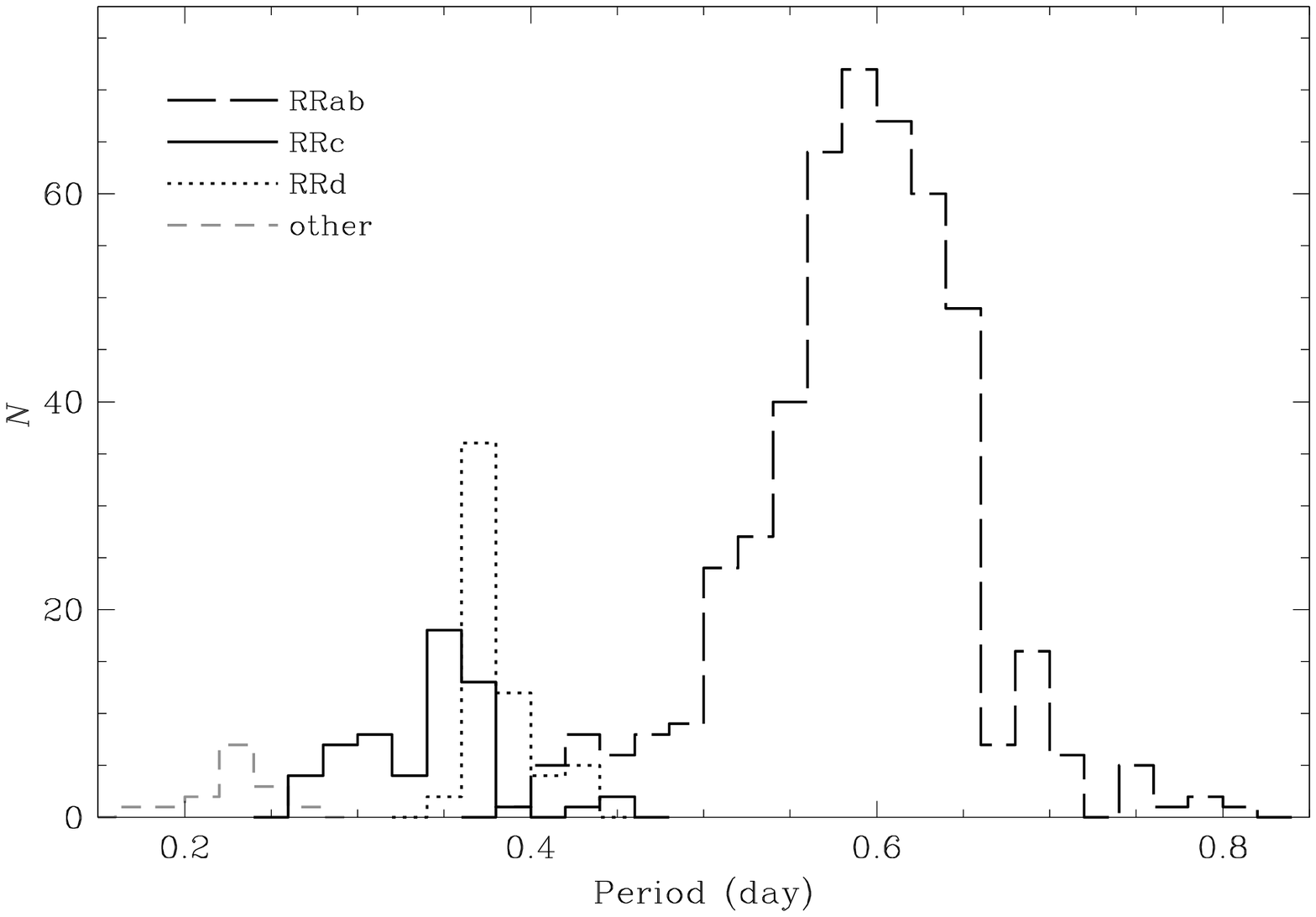}}
\FigCap{Distribution of periods for RR~Lyr stars from the SMC. The bins are 
0.02 days wide. Long-dashed line represents period distribution of RRab stars, 
solid line -- RRc stars, dotted line -- RRd stars (first overtone periods), 
and short-dashed -- other stars.} 
\end{figure} 
Period distribution for the global sample is shown in Fig.~3. The mean period 
of ab-type RR~Lyr stars, and simultaneously the most preferred period, is 
${\langle P_{ab}\rangle=0.589}$ days. This result is in the ideal agreement 
with Graham's (1975) determination of the mean period of RRab variables near 
NGC121. ${\langle P_{ab}\rangle}$ is believed to be an indicator of the mean 
metallicity -- it increases with decreasing metallicity. However, one should 
remember that Clement \etal (2001), in a~study of the properties of the RR~Lyr 
stars in Galactic globular clusters, suggested that the correlation breaks 
down when the mean period is larger than 0.60~days. 

\renewcommand{\arraystretch}{1}
\renewcommand{\TableFont}{\footnotesize}
\setcounter{table}{5}
\MakeTable{
l@{\hspace{10pt}}
l@{\hspace{4pt}}
l}{12.5cm}{Cross-identification of stars detected in the overlapping regions}
{\hline
\noalign{\vskip3pt}
\multicolumn{1}{c}{Star ID} & \multicolumn{2}{c}{Fields}\\
\noalign{\vskip3pt}
\hline
\noalign{\vskip3pt}
OGLE004227.44$-$733655.1 & SMC$\_$SC2 & SMC$\_$SC3\\
OGLE004222.96$-$733223.8 & SMC$\_$SC2 & SMC$\_$SC3\\
OGLE004225.29$-$730349.8 & SMC$\_$SC2 & SMC$\_$SC3\\
OGLE004533.91$-$733240.2 & SMC$\_$SC3 & SMC$\_$SC4\\
OGLE004526.76$-$732801.7 & SMC$\_$SC3 & SMC$\_$SC4\\
OGLE004531.73$-$732634.2 & SMC$\_$SC3 & SMC$\_$SC4\\
OGLE004533.00$-$731255.1 & SMC$\_$SC3 & SMC$\_$SC4\\
OGLE004533.33$-$730120.2 & SMC$\_$SC3 & SMC$\_$SC4\\
OGLE004528.73$-$725109.7 & SMC$\_$SC3 & SMC$\_$SC4\\
OGLE004827.91$-$733108.7 & SMC$\_$SC4 & SMC$\_$SC5\\
OGLE004828.10$-$731442.2 & SMC$\_$SC4 & SMC$\_$SC5\\
OGLE004833.02$-$731800.0 & SMC$\_$SC4 & SMC$\_$SC5\\
OGLE004835.73$-$731330.1 & SMC$\_$SC4 & SMC$\_$SC5\\
OGLE005126.54$-$725205.3 & SMC$\_$SC5 & SMC$\_$SC6\\
OGLE005135.18$-$724438.1 & SMC$\_$SC5 & SMC$\_$SC6\\
OGLE005428.43$-$724840.3 & SMC$\_$SC6 & SMC$\_$SC7\\
OGLE005430.51$-$724240.9 & SMC$\_$SC6 & SMC$\_$SC7\\
OGLE005723.39$-$730159.6 & SMC$\_$SC7 & SMC$\_$SC8\\
OGLE005725.77$-$725438.5 & SMC$\_$SC7 & SMC$\_$SC8\\
OGLE005728.63$-$724135.0 & SMC$\_$SC7 & SMC$\_$SC8\\
OGLE005728.85$-$723454.6 & SMC$\_$SC7 & SMC$\_$SC8\\
OGLE010029.58$-$725454.3 & SMC$\_$SC8 & SMC$\_$SC9\\
OGLE010316.37$-$724816.2 & SMC$\_$SC8 & SMC$\_$SC9\\
OGLE010326.63$-$721409.8 & SMC$\_$SC9 & SMC$\_$SC10\\
OGLE010324.09$-$720545.7 & SMC$\_$SC9 & SMC$\_$SC10\\
\noalign{\vskip3pt}
\hline}

The period distribution of the first overtone pulsators indicates two 
preferred periods. Most of the RRc stars prefer periods near ${\langle P_c 
\rangle=0.357}$~days, but there is also a~significant excess of stars around 
${P=0.29}$~days. Such short-period RR~Lyr stars also exist in other 
environments, \eg in the LMC (Alcock \etal 1996) and in the globular cluster 
IC~4499 (Walker and Nemec 1996). It has been suggested (van Albada and Baker 
1973) that these stars are second overtone (RRe) pulsators. The pulsation of 
RR~Lyr stars in overtone modes was theoretically analyzed by Stellingwerf 
\etal (1987). 

RRd variable stars (first-overtone periods) and RRc stars occupy different, 
but overlapping regions in the period distribution histogram. Periods of the 
first overtone in the double-mode RR~Lyr stars range from about 0.354 to 0.435 
days with the mean period ${\langle P_d\rangle=0.382}$ days. Single-mode 
first-overtone pulsators have typically shorter or longer periods than 
double-mode RR~Lyr stars (we found three candidates for RRc stars with periods 
between 0.428~and 0.446~days). 

\Subsection{Intensity Mean Photometry}
{\it BVI} intensity mean photometry of each object from our sample of RR~Lyr 
candidates was derived by integrating the light curve converted to intensity 
units. The light curve was approximated by the Fourier series of fifth order 
and results were converted back to the magnitude scale. Accuracy of the mean 
{\it I}-band photometry is about 0.02~mag while that of the {\it V}-band and 
{\it B}-band about 0.03~mag and 0.08~mag, respectively, what is a consequence 
of smaller number of observations in the {\it BV}-band filters. 

To obtain the mean brightness of the SMC RR~Lyr stars, we prepared histograms 
of the {\it BVI}-band apparent magnitudes separately for RRab, RRc and RRd 
variable stars. In the next step we fitted a~Gaussian to each histogram, and 
obtained the most preferred brightness of these stars. Then we repeated this 
procedure using magnitudes corrected for interstellar extinction. Results are 
presented in Tables~7 and~8. 
\setcounter{table}{6}
\MakeTable{lccccccccc}{12.5cm}{Mean $I$, $V$ and $B$ magnitudes of RR~Lyr stars}
{\hline
\multicolumn{1}{l}{Type}& $I$ & $\sigma_I$ & $V$ & $\sigma_V$ & $B$ & $\sigma_B$\\
\hline
RRab & $19.14\pm0.01$ & $0.20$ & $19.74\pm0.01$ & $0.23$ &
$20.10\pm0.01$ & $0.25$\\ RRc & $19.27\pm0.01$ & $0.15$ &
$19.73\pm0.01$ & $0.15$ & $20.00\pm0.01$ & $0.23$\\ RRd &
$19.17\pm0.01$ & $0.17$ & $19.69\pm0.01$ & $0.20$ &
$19.94\pm0.02$ & $0.22$\\
\hline}
\MakeTable{lccccccccc}{12.5cm}{Mean $I$, $V$ and $B$ extinction 
free magnitudes of RR~Lyr stars}
{\hline
\multicolumn{1}{l}{Type}& $I_0$ & $\sigma_{I_0}$ & $V_0$ & $\sigma_{V_0}$ & $B_0$ & $\sigma_{B_0}$\\
\hline
RRab & $18.97\pm0.01$ & $0.20$ & $19.45\pm0.01$ & $0.23$ &
$19.73\pm0.01$ & $0.25$\\ RRc & $19.10\pm0.01$ & $0.15$ &
$19.45\pm0.01$ & $0.15$ & $19.63\pm0.02$ & $0.24$\\ RRd &
$19.00\pm0.01$ & $0.17$ & $19.42\pm0.01$ & $0.20$ &
$19.58\pm0.02$ & $0.24$\\
\hline}

\Subsection{Period-Amplitude Relation for RR~Lyr Stars}
We determined amplitudes of light curves by fitting Fourier series of fifth 
order and calculating difference between minimal and maximal values of the 
function. The period-amplitude diagram for the SMC RR~Lyr stars is shown in 
Fig.~4. The symbols are as follows: black dots indicate the position of RRab 
and RRc stars, triangles and diamonds represent RRd stars (first-overtone and 
fundamental mode, respectively) and crosses mark other stars. 

\begin{figure}[p]
\centerline{\includegraphics[bb=20 30 570 750, width=11.6cm]{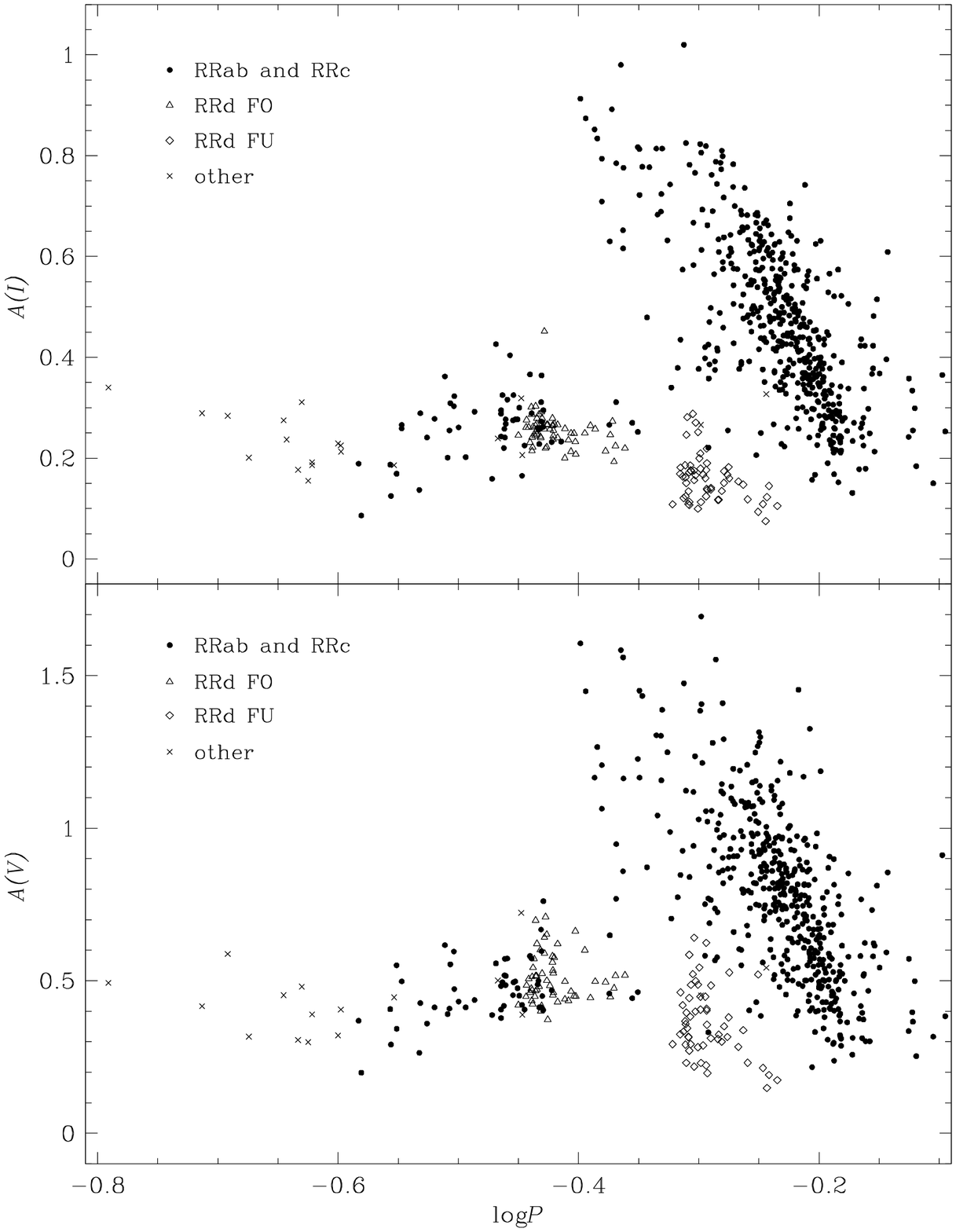}}
\FigCap{Period--{\it I}-amplitude (upper panel) and period--{\it V}-amplitude 
(lower panel) diagrams for RR~Lyr stars from the SMC. Small dots represent
RRab and RRc stars, triangles and diamonds represent RRd stars
(first-overtone and fundamental mode, respectively), and crosses mark other
stars.} 
\end{figure} 

Well-known dependence of the amplitudes on periods for RRab stars is seen, 
although some stars, especially RRab with shorter periods, do not follow this 
relation. The first overtone RR~Lyr variables show weak correlation of 
amplitudes and periods but, inversely to the fundamental mode pulsators, 
amplitude slightly increases with increasing periods. 

Comparison of amplitudes of fundamental mode and first overtone of RRd stars 
confirms that the dominant mode of pulsation in the majority of double-mode 
RR~Lyr stars is the first overtone. The mean ratio of the first-harmonic and 
fundamental mode amplitudes is about 1.6. 

\Subsection{Fourier Parameters of Light Curve Decomposition}
Fourier decomposition of light curves of RR~Lyr stars has been widely used for 
analyzing their properties. Several empirical and theoretical relations 
between Fourier parameters and the physical parameters of RR~Lyr variables 
have been proposed (\eg Kov{\'a}cs and Walker 2001). 

\begin{figure}[p]
\centerline{\includegraphics[bb=30 50 570 750, width=11.6cm]{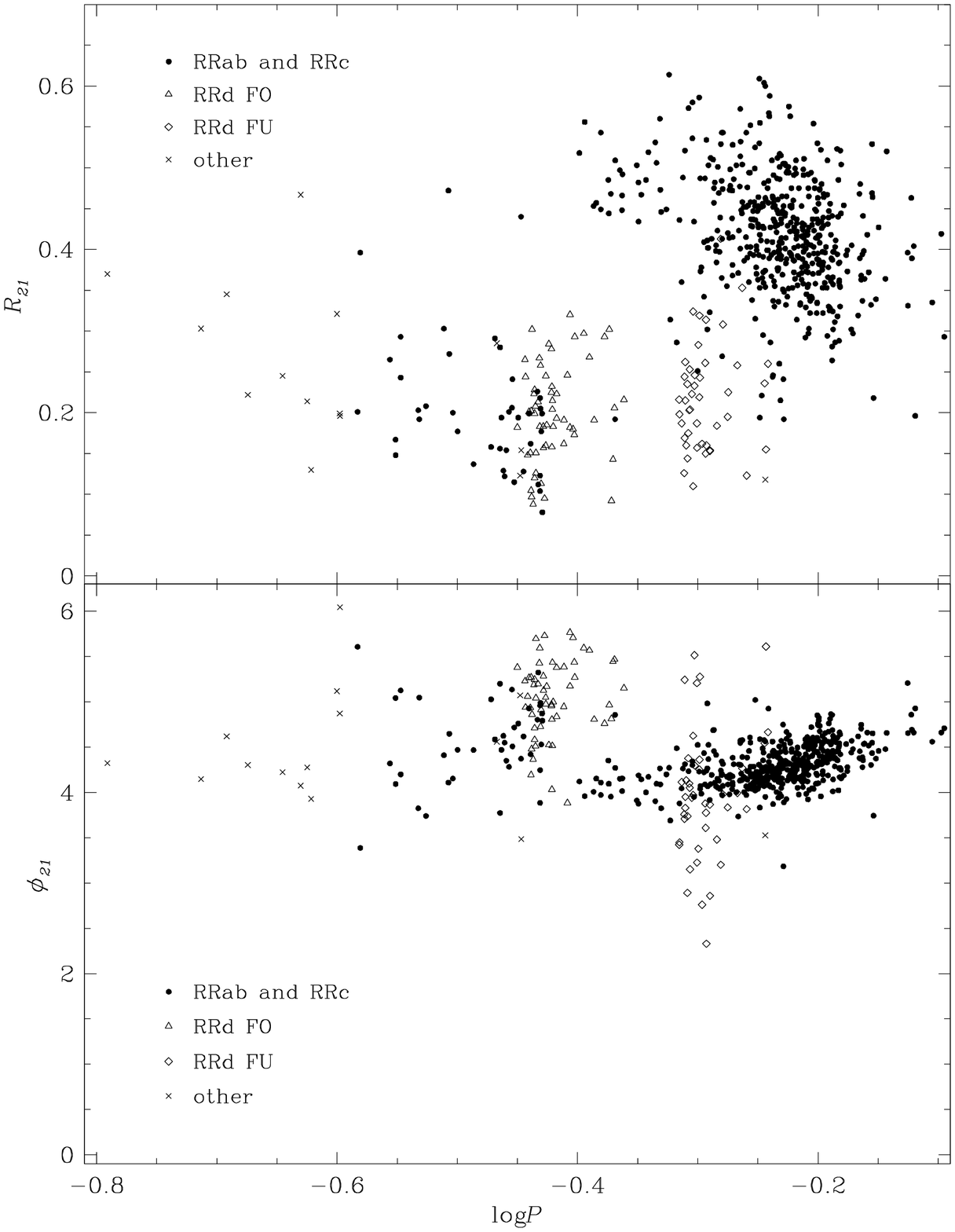}}
\FigCap{$R_{21}$ and $\phi_{21}$ \vs $\log{P}$ diagrams for RR~Lyr stars 
from the SMC. Small dots represent RRab and RRc stars, triangles and diamonds 
represent RRd stars (first-overtone and fundamental mode, respectively), and 
crosses mark other stars.} 
\end{figure} 
Fifth order Fourier series were fitted to the magnitude scale {\it I}-band 
light curve. Then, we derived $R_{21}$, $R_{31}$, $\phi_{21}$ and $\phi_{31}$ 
Fourier parameters, where $R_{ij}=A_i/A_j$, $\phi_{ij}=\phi_{i}-i \phi_{j}$. 
$A_i$ and $\phi_{i}$ are the amplitudes and phases of ($i-1$) harmonic of the 
Fourier decomposition of light curve. 

Light curves of double-mode pulsators were decomposed to the sum of two 
Fourier series of fourth order corresponding to both periodicities. Then we 
calculated Fourier parameters for both pulsating modes. 

In Fig.~5 we present $R_{21}-\log{P}$ and $\phi_{21}-\log{P}$ diagrams 
constructed for our sample of RR~Lyr stars. The first-overtone and fundamental 
mode pulsators are well separated in both diagrams. $R_{21}$ for RRab stars 
tends to become smaller as the periods become longer, what means that the 
light curves of long-period ab-type RR~Lyr are more sinusoidal than the light 
curves of short-period RRab stars. 

Fourier parameters of the first-overtone mode of RRd stars fall in the 
sequence of the single-mode first overtone RR~Lyr stars, but the fundamental 
mode pulsation have $R_{21}$ smaller than corresponding single-mode RR~Lyr 
variables. This means that fundamental-mode pulsations in RRd stars have not 
only smaller amplitude, but the light curves are more sinusoidal than the 
light curves of RRab stars. 

\Subsection{Petersen Diagram}
RR~Lyr stars pulsating simultaneously in the fundamental mode and the first 
overtone are especially interesting as a test of structural and evolutionary 
models of the horizontal branch stars. A powerful tool for diagnosing stellar 
models and for determination of masses, absolute magnitudes and metallicities 
of stars is the so called Petersen diagram, in which the period ratio of two 
excited modes is plotted against the longer period. Position of stars in the 
Peterson diagram can be precisely determined, because periods are the 
observables that are measured with the highest precision. 

\begin{figure}[htb]
\centerline{\includegraphics[bb=20 220 550 720, width=11cm]{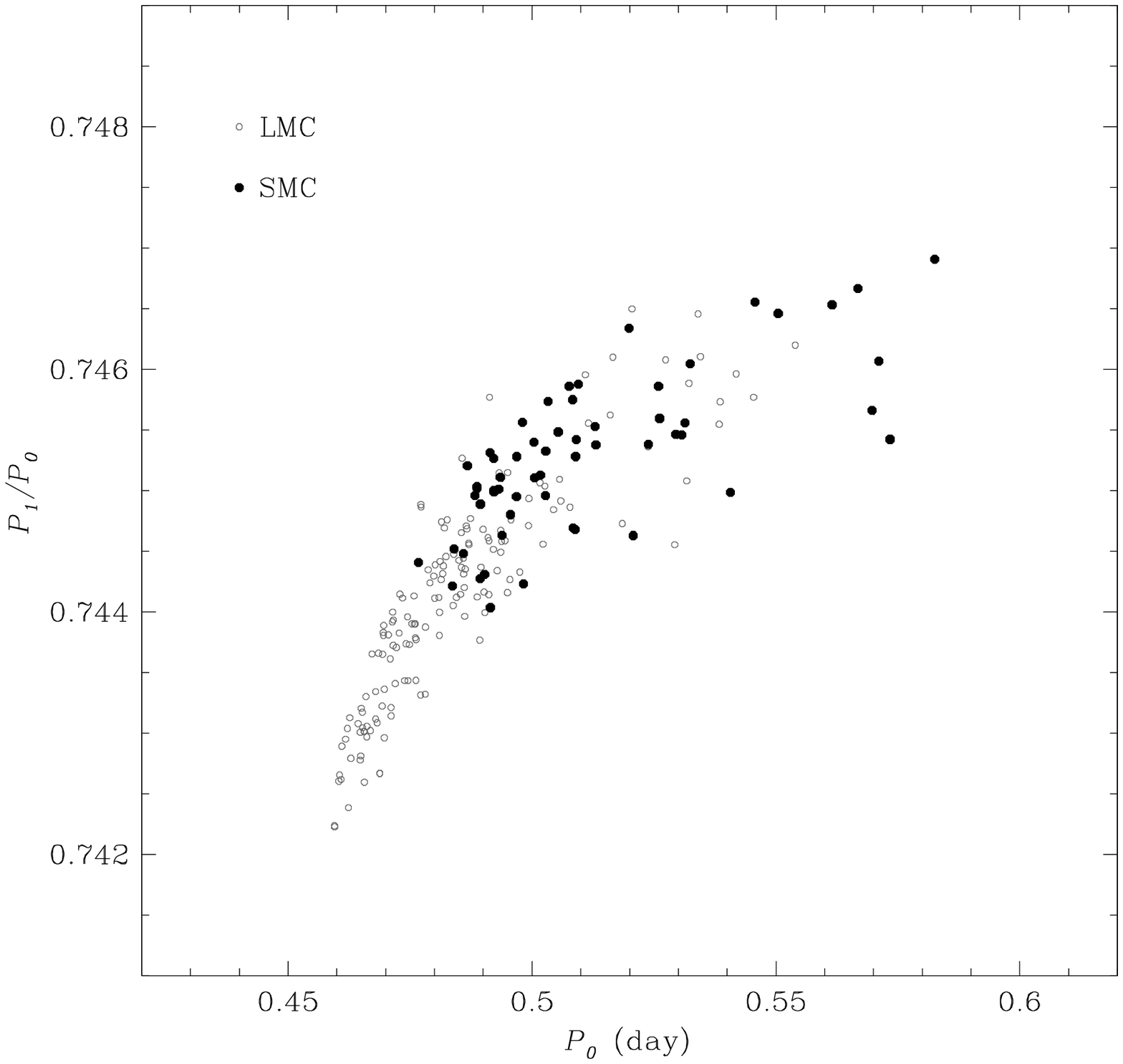}}
\FigCap{Petersen diagram for the LMC and SMC double-mode RR~Lyr stars. The 
open circles mark RRd stars from the LMC, the filled circles represent RRd 
stars from the SMC.} 
\end{figure} 
In Fig.~6 we present Petersen diagram of all discovered RRd stars. For 
comparison, we also plotted in the same diagram double mode RR~Lyr stars from 
the LMC (catalog of RR~Lyr stars from the LMC will be published in the 
forthcoming paper). 

The SMC and LMC double-mode RR~Lyr stars evidently form the same relation 
$P_1/P_0$ \vs $P_0$, but RRd stars from the SMC prefer longer periods and 
larger period ratios than the LMC RRd stars. Pulsation models suggest that the 
position of a given RRd star in the Petersen diagram is a function of stellar 
mass and metallicity. We can expect that the shift in the Petersen diagram 
between RRd stars from the LMC and the SMC is caused by different mean 
metallicity of RR~Lyr stars in both galaxies. More information about analysis 
of the Peterson diagrams can be found in Popielski, Dziembowski and Cassisi 
(2000). 

\Section{Completeness of the Catalog}
We estimated completeness of the Catalog of RR~Lyr stars in similar manner as 
for the catalog of Cepheids from the Small Magellanic Cloud (Udalski \etal 
1999), \ie by comparison of objects located in the overlapping regions of the 
adjacent fields. Ten such regions exist between our fields (Fig.~1) allowing 
to perform 20 tests of paring objects from a~given and neighboring fields. In 
total, 57 stars from Tables 2--5 should be theoretically paired with 
counterparts in the overlapping fields. We found counterparts in 50 cases. 
Counterparts of four from seven unpaired objects had the number of 
observations smaller than 50 and thus they were not searched for variability. 
Smaller number of measurement points in the edge regions of the fields is due 
to non-perfect pointing of the telescope. The remaining three objects were 
missed because of severe blending with another stars or because of unusually 
small amplitude of variability. 

Udalski \etal (1998a) estimated completeness of detection of stars in the SMC 
OGLE fields using artificial star tests. For stars as bright as RR~Lyr the 
completeness was found to be between 90 and 95\%, depending on the field 
density. In our sample completeness should be much higher, because the stars 
are being detected on the DIA reference images, obtained by co-adding 20 best 
frames for each field. We conclude that the completeness of our sample should 
be about 90\%. 

\Section{Conclusions}
The long term photometric surveys of dense stellar systems may provide large 
samples of all types of variable stars (Paczy{\'n}ski 2000). In this paper we 
present the catalog of RR~Lyr variable stars -- objects playing a major role 
in studies of the processes of stellar pulsation, post-main-sequence evolution 
of stars, calibration of extragalactic distances, and structure and age of 
galaxies. This is the largest sample of RR~Lyr variables detected in the SMC 
so far. 

We reported detection of 57 double-mode RR~Lyr stars. We also found RR~Lyr 
variables exhibiting two closely spaced frequencies, most probably related to 
non-radial pulsations. 

About 10\% of the detected RR Lyr variables are RRc stars. Next 10\% of the 
\renewcommand{\TableFont}{\footnotesize}
\setcounter{table}{8}
\MakeTable{
l@{\hspace{3pt}}
l@{\hspace{6pt}}
l@{\hspace{6pt}}
l@{\hspace{4pt}}
c@{\hspace{4pt}}
c@{\hspace{4pt}}
}{12.5cm}{RR Lyr stars close to the SMC clusters}
{\hline
\noalign{\vskip3pt}
\multicolumn{1}{c}{Field} & \multicolumn{1}{c}{Star ID} & 
Cluster & Alternative & Cluster & Distance from \\
 & & name & cluster & radius & the center \\
 & & OGLE-CL- & name & [\arcs] & [\arcs]  \\
\noalign{\vskip3pt}
\hline
\noalign{\vskip3pt}
 SMC$\_$SC2 & OGLE004222.96$-$733223.8 & SMC0014 & & ~24 & 33 \\
 SMC$\_$SC3 & OGLE004504.31$-$725623.9 & SMC0024 & & ~57 & 68 \\
 SMC$\_$SC4 & OGLE004526.76$-$732801.7 & SMC0027 & B39 & ~36 & 51 \\
 SMC$\_$SC4 & OGLE004610.66$-$730355.6 & SMC0183 & & ~10 &  ~~1 \\
 SMC$\_$SC4 & OGLE004817.88$-$731815.2 & SMC0048 & B47 & ~49 & 67 \\
 SMC$\_$SC5 & OGLE004833.02$-$731800.0 & SMC0048 & B47 & ~49 & 25 \\
 SMC$\_$SC5 & OGLE004907.44$-$730617.5 & SMC0052 & H86-104 & ~18 & 25 \\
 SMC$\_$SC5 & OGLE004913.17$-$730651.0 & SMC0052 & H86-104 & ~18 & 21 \\
 SMC$\_$SC5 & OGLE004922.63$-$731220.9 & SMC0053 & & ~30 & 30 \\
 SMC$\_$SC5 & OGLE004924.05$-$732231.3 & SMC0054 & L39 & ~27 & 30 \\
 SMC$\_$SC5 & OGLE004924.31$-$731447.2 & SMC0195 & BS42 & ~28 & 35 \\
 SMC$\_$SC5 & OGLE004954.79$-$730321.3 & SMC0057 & B52 & ~49 & 67 \\
 SMC$\_$SC5 & OGLE004936.31$-$725229.2 & SMC0058 & H86-109 & ~36 & 51 \\
 SMC$\_$SC5 & OGLE005120.92$-$730920.1 & SMC0069 & NGC290 & ~36 & 36 \\
 SMC$\_$SC6 & OGLE005202.85$-$725707.3 & SMC0074 & K29,L44 & ~42 & 44 \\
 SMC$\_$SC6 & OGLE005243.99$-$724740.5 & SMC0085 & B66 & ~25 & 17 \\
 SMC$\_$SC7 & OGLE005549.04$-$725335.1 & SMC0105 & H86-165 & ~44 & 54 \\
 SMC$\_$SC7 & OGLE005612.31$-$731222.5 & SMC0106 & & ~26 & 15 \\
 SMC$\_$SC8 & OGLE005820.37$-$723841.4 & SMC0113 & L61-331 & ~24 & 19 \\
SMC$\_$SC11 & OGLE010728.50$-$724527.6 & SMC0156 & L80 & ~41 & 42 \\ 
SMC$\_$SC11 & OGLE010836.97$-$725321.6 & SMC0159 & NGC419 &102 & 80 \\
\noalign{\vskip3pt}
\hline}
whole sample are RRd variables. Relatively small number of the first-overtone 
pulsators and the value of the average period of RRab stars (0.589 days) 
places the SMC RR~Lyr stars near the Oosterhoff type~I group. 

Additionally, we conducted a search of RR~Lyr stars in the SMC clusters. 
Discovery of RR~Lyr variables in the clusters younger than NGC121 (hosting 
four RR~Lyr stars), would have a significant consequences on our understanding 
of evolution of old, solar-mass stars. We used the OGLE catalog of the SMC 
clusters (Pietrzy{\'n}ski \etal 1998) containing 238 clusters, their 
coordinates and angular sizes. We scanned our sample of RR~Lyr variables 
looking for stars in a distance smaller than 1.5 cluster radius from the 
cluster center. We obtained the list of 21 RR~Lyr stars located close to the 
line-of-sight to, in total, 19 SMC clusters. Only in two cases 
(OGLE-CL-SMC0048 and OGLE-CL-SMC0052) we detected two RR~Lyr stars in the 
cluster neighborhood. 
It is very probable that the majority of these cases are just  optical 
coincidences, because more or less similar number of stars could be expected 
in the background of SMC clusters (comparing the area of the clusters to the 
area of all our fields). Nevertheless, because of potential importance of the 
discovery of RR~Lyr variables in the SMC clusters, we provide in Table~9 the 
list of stars located close to the SMC clusters. 

\Acknow{We would like to thank Prof.~Bohdan Paczy{\'n}ski for many discussions 
and valuable suggestions. We are very grateful to Prof.~Wojciech Dziembowski 
for helping us in the classification of stars. The paper was partly supported 
by the Polish KBN grant BST to Warsaw University Observatory. Partial support 
for the OGLE   project was provided with the NSF grants AST-9820314 and 
AST-0204908 and NASA grant NAG5-12212 to B.~Paczy\'nski. We acknowledge usage 
of the Digitized Sky Survey which was produced at the Space Telescope Science 
Institute based on photographic data obtained using the UK Schmidt Telescope, 
operated by the Royal Observatory Edinburgh.}

\end{document}